\documentstyle [12pt,a4,epsf] {article}

\newcommand{\half} {\frac{1}{2}}
\newcommand{\e} { {\rm e} }
\newcommand{\ie} {{\it i.e.}}
\newcommand{\eg} {{\it e.g.}}
\newcommand{\be} {\begin{equation}}
\newcommand{\ee} {\end{equation}}
\newcommand{\bea} {\begin{eqnarray}}
\newcommand{\eea} {\end{eqnarray}}
\newcommand{\dd} {{\rm d}}

\begin{document}


\title {Adsorbed and Grafted \\
            Polymers at Equilibrium}
\author {Roland R. Netz \\
         Max-Planck Institute for Colloids and Interfaces\\
         D-14424 Potsdam, Germany \\
\and
David Andelman \\
         School of Physics and Astronomy   \\
         Raymond and Beverly Sackler Faculty of Exact Sciences \\
         Tel Aviv University, Ramat Aviv, Tel Aviv 69978, Israel \\
         \\}
\date{January 2000}
\maketitle

\begin{abstract}
\setlength {\baselineskip} {2pt}

This chapter deals with various aspects related to the adsorption
of long chain-like macromolecules (polymers) onto solid surfaces.
Physical aspects of the adsorption mechanism are elaborated mainly
at thermodynamical equilibrium. The basic features needed in
modeling of this adsorption are discussed. Among other aspects, we
address the type of polymer/surface interaction, the solvent
quality,  and the characteristics of the surface. We first discuss
the adsorption of a single and long polymer chain to the surface.
The surface interaction is modeled by a potential well with a
long-ranged attractive tail. A simple mean--field theory
description is presented and the concept of polymer ``blobs" is
used to describe the conformation of the chain at the surface. The
thickness of the adsorbed layer depends on several polymer and
surface parameters. Fluctuation corrections to mean--field theory
are also discussed. We then review adsorption as well as depletion
processes in the many-chain case. Profiles, changes in the surface
tension and polymer surface excess are calculated within 
mean--field theory. Corrections 
due to fluctuations in good solvent are
taken into account using scaling concepts. The proximal exponent
is introduced in analogy to surface critical phenomena. The
interaction between two surfaces with adsorbed polymer layers is
discussed, and various cases leading to attractive and repulsive
interactions are mentioned. Polyelectrolytes are of practical
importance due to their water solubility; we give a short summary
of recent progress in this rapidly evolving field.  The behavior
of grafted polymers, \ie, polymers which are anchored with one end
to a solid substrate is also reviewed. Here  we discuss the shape
of the density profile, the osmotic pressure upon lateral
compression of the brush, and the force of interaction between two
surfaces each being coated with a grafted polymer layer. The
latter case is important in understanding colloidal stability.
\end{abstract}

 \pagebreak \setlength
{\baselineskip} {2pt}

\section{Introduction}
\setcounter{equation}{0}

In this chapter, we review the basic mechanisms underlying
adsorption of long chain molecules on solid surfaces such as
oxides. We concentrate on the physical aspects of adsorption and
summarize the main theories which have been proposed. This chapter
should be viewed as a general introduction to the problem of
polymer adsorption at thermodynamical equilibrium.  For a
selection of previous review articles see
Refs.~\cite{cscv86}-\cite{fl97}, while more detailed treatments
are presented in  two books on this subject,
Refs.~\cite{fleer,erich}. We do not attempt to explain any specific
polymer/oxide system and do not emphasize experimental results and
techniques. Rather, we detail how concepts taken from statistical
thermodynamics and interfacial science can explain general and
{\em universal} features of polymer adsorption. The present
chapter deals with equilibrium properties whereas the following
chapter by Cohen Stuart and de Keizer is about kinetics. We first
outline the basic concepts and assumptions that are
employed throughout the chapter.

\subsection{Types of Polymers}

The polymers considered here are taken as linear and long chains,
such as schematically depicted in Fig.~1a. We do not address the
more complicated case of branched polymers at interfaces, although
a considerable amount of work has been done on such 
systems~\cite{Grosberg}.
In Fig.~1b we schematically present an example of a branched
polymer. Moreover, we examine mainly homopolymers where the
polymers are composed of the same repeated unit (monomer). We
discuss separately, in Sect.~8, extensions to adsorption of block
copolymers and to polymers that are terminally grafted to the
surface on one side (``polymer brushes"). In most of this review
we shall assume that the chains are neutral. The charged case,
\ie, where each or a certain fraction of monomers carries an
electric charge, as depicted in Fig.~1c,
 is still not very well understood and depends on
additional parameters such as the surface charge density, the
polymer charge, and the ionic strength of the solution. We address
shortly adsorption of polyelectrolytes in Sect. \ref{sectionpe}.
Furthermore, the chains are considered to be flexible. The
statistical thermodynamics of flexible chains is rather well
developed and the theoretical concepts can be applied with a
considerable degree of confidence. Their large number of
conformations play a crucial role in the adsorption, causing a
rather {\em diffusive}
layer extending away from the surface into the solution.
This is in contrast to rigid chains, which usually form dense adsorption
layers on surfaces.

\subsection{Solvent Conditions}

Polymers in solution can experience three types of solvent
conditions. The solvent is called ``good'' when the
monomer-solvent interaction is more favorable than the
monomer-monomer one. Single polymer chains in good solvents have
``swollen'' spatial configurations, reflecting the effective
repulsion between monomers. In the opposite case of ``bad''
(sometimes called ``poor") solvent conditions, the effective
interaction between monomers is attractive, leading to collapse of
the chains and to their precipitation from the solution (phase
separation between the polymer and the solvent). In the third and
intermediate solvent condition, called ``theta'' solvent, the
monomer-solvent and monomer-monomer interactions are equal in
strength.  The chains are still soluble, but their spatial
configurations and solution properties differ from the
good-solvent case.

The theoretical concepts and methods leading to these three
classes make up a large and central part of polymer physics and
are summarized in text
books~\cite{Grosberg}-\cite{Cloizeaux}.
In general, the solvent quality depends mainly on the specific
chemistry determining the interaction between the solvent
molecules and monomers. It also can be changed by varying the
temperature.

\subsection{Adsorption and Depletion}

Polymers can adsorb spontaneously from solution onto surfaces if
the interaction between the polymer and the surface is more
favorable than that of the solvent with the surface. For example,
a polymer like polyethylene oxide (PEO) is soluble in water but
will adsorb on various hydrophobic surfaces and on the water/air
interface.  This is the case of equilibrium adsorption where the
concentration of the polymer monomers  increases close to the
surface with respect to their concentration in the bulk solution.
We discuss this phenomenon at length both on the level of a single
polymer chain (valid only for extremely dilute polymer solutions),
see Sect. 2, and for polymers adsorbing from (semi-dilute)
solutions, see Sect. 3. In Fig.~2a we schematically show the
volume fraction profile $\phi(z)$ of monomers as a function of the
distance $z$ from the adsorbing substrate. In the bulk, that is
far away from the substrate surface, the volume fraction of the
monomers is $\phi_b$, whereas at the surface, the corresponding
value is $\phi_s > \phi_b$.
The theoretical models address questions in relation to the
polymer conformations at the interface, the local concentration of
polymer in the vicinity of the surface and the total amount of
adsorbing polymer chains. In turn, the knowledge of the polymer
interfacial behavior is used to calculate thermodynamical
properties like the surface tension in the presence of polymer
adsorption.

 The opposite case of {\em
depletion} occurs when the monomer-surface interaction is less
favorable than the solvent-surface interaction. This is, \eg, the
case for polystyrene in toluene which is depleted from a mica
substrate. The depletion layer is defined as the layer adjacent to
the surface from which the polymers are depleted. Their
concentration in the vicinity of the surface is lower than the
bulk value, as shown schematically in Fig.~2b.

\subsection{Surface--Polymer Interactions}

Equilibrium adsorption of polymers is only one of the methods used
to create a change in the polymer concentration close to a
surface. For an adsorbed polymer, it is interesting to look at the
detailed conformation of a single polymer chain at the substrate.
One distinguishes sections of the polymer which are bound to the
surface, so-called trains, sections which form loops, and the end
sections of the polymer chain, which can form dangling tails. This
is schematically depicted in Fig.~3a. Two other methods to produce
polymer layers at surfaces are commonly used for polymers which do
not spontaneously adsorb on a given surface.

(i) In the first method, the polymer is chemically attached
(grafted) to the surface by one of the chain ends, as shown in
Fig.~3b. In good solvent conditions the polymer chains look like
``mushrooms'' on the surface when the distance between grafting
points is larger than the typical size of the chains. In some
cases, it is possible to induce a much higher density of the
grafting resulting in a polymer ``brush'' extending in the
perpendicular direction from the surface, as is discussed in
detail in Sect. \ref{sectionbrush}.

(ii) A variant on the grafting method is to use a diblock
copolymer made out of two distinct blocks, as shown in Fig.~3c.
The first block is insoluble and attracted to the substrate,
 and thus acts as an ``anchor'' fixing
the chain to the surface; it is drawn as a thick line in Fig.~3c.
It should be long enough to cause irreversible fixation on the
surface. The other block is a soluble one (the ``buoy''), forming
the brush layer. For example, fixation on hydrophobic surfaces
from a water solution can be made using a polystyrene-polyethylene
oxide (PS-PEO) diblock copolymer. The PS block is insoluble in
water and attracted towards the substrate, whereas the PEO forms
the brush layer. The process of diblock copolymer fixation has a
complex dynamics during the formation stage but is very useful in
applications~\cite{napper}.

\subsection{Surface Characteristics}

Up to now we outlined the polymer properties. What about the
surface itself? Clearly, any adsorption process will be sensitive
to the type of surface and its internal structure. As a starting
point we assume that the solid surface is atomically smooth, flat,
and homogeneous, as shown in Fig.~4a. This ideal solid surface is
impenetrable to the chains and imposes on them a surface
interaction. The surface potential can be either short-ranged,
affecting only the monomers which are in direct contact with the
substrate or in close vicinity of the surface. The surface can
also have a longer range effect, like van der Waals, or
electrostatic interactions, if it is charged. Interesting
extensions beyond ideal surface conditions are expected in several
cases: (i)~rough or corrugated surfaces, such as depicted in
Fig.~4b;   (ii)~surfaces that are curved, \eg, adsorption on
spherical colloidal particles, see Fig.~4c; (iii)~substrates which
are chemically inhomogeneous, \ie, which show some lateral
organization, as shown schematically in Fig.~4d; (iv)~surfaces
that have internal degrees of freedom like surfactant monolayers;
and (v)~polymer adsorbing on ``soft" and ``flexible" interfaces
between two immiscible fluids or at the liquid/air surface. We
briefly mention those situations in Sects. 5-7.

\subsection{Polymer Physics}

Before turning to the problem of polymer adsorption let us briefly
mention some  basic principles of polymer theory. For a more
detailed exposure the reader should consult the books by Flory, de
Gennes, or  Des Cloizeaux and
Jannink~\cite{Flory1,degennes,Cloizeaux}. The main parameters
needed to describe a flexible polymer chain are the polymerization
index $N$, which counts the number of repeat units or monomers,
and the Kuhn length $a$, which corresponds to the spatial size of
one monomer or the distance between two neighboring monomers. The
monomer size ranges from $1.5$ \AA, as for example for
polyethylene, to a few nanometers for biopolymers~\cite{Flory1}.
In contrast to other molecules or particles, a polymer chain
contains not only translational and rotational degrees of freedom,
but also a vast number of conformational degrees of freedom. For
typical polymers, different conformations are produced by
torsional rotations of the polymer backbone bonds as shown
schematically in Fig.~5a for a polymer consisting of four bonds of
length $a$ each. A satisfactory description of flexible chain
conformations is achieved with the (bare) statistical weight for a
polymer consisting of $N+1$ monomers

\be \label{Wiener}
{\cal P}_N = \exp\left\{ -\frac{3}{2 a^2}
\sum_{i=1}^{N} ({\bf r}_{i+1}-{\bf r}_i)^2 \right\}
\ee

\noindent which assures that each bond vector, given by ${\bf
r}_{i+1}-{\bf r}_i$ with $i=1, \ldots , N$, treated for
convenience as a fluctuating Gaussian variable, has a mean length
given by the Kuhn length, \ie,

\[ \langle ({\bf r}_i-{\bf r}_{i+1})^2 \rangle = a^2\]

\noindent In most theoretical approaches, it is useful to take the
simplification one step further and represent the polymer as a
continuous line, as shown in Fig.~5b,  with the statistical weight
for each conformation given by Eq. (\ref{Wiener}) in the continuum
limit.
The Kuhn length $a$ in this limit loses its geometric
interpretation as the monomer size, and simply becomes an elastic
parameter which is tuned such as to ensure the proper behavior of
the large-scale properties of this continuous line, as is detailed
below. Additional effects, such as a local bending rigidity,
preferred bending angles (as relevant for trans-gauche isomery
encountered for saturated carbon backbones), and hindered
rotations can be taken into account by defining an effective
polymerization index and an effective Kuhn length. In that sense,
we always talk about effective parameters $N$ and $a$, without
saying so explicitly. Clearly, the total polymer length in the
completely extended configuration is $L = aN$. However, the
average spatial extent of a polymer chain in solution is typically
much smaller. An important quantity characterizing the size of a
polymer coil  is the average end-to-end radius $R_e$. For the
simple Gaussian polymer model defined above, we obtain

\be
R_e^2 = \langle ( {\bf r}_{N+1} - {\bf r}_1)^2 \rangle = a^2 N \ee

\noindent In a more general way, one describes the scaling
behavior of the end-to-end radius for large values of $N$ as $R_e
\sim a N^{\nu}$. For an ideal polymer chain, \ie, for a polymer
whose individual monomers do not interact with each other, the
above result implies $\nu = 1/2$. This result holds only for
polymers where the attraction between monomers (as compared with
the monomer-solvent interaction) cancels the steric repulsion due
to the impenetrability of monomers. This situation can be achieved
in special solvent conditions called  ``theta" solvent as was
mentioned above. In a theta solvent, the polymer chain is not as
swollen as in good solvents but is not collapsed on itself either,
as it is under bad solvent conditions.

For good solvents, the steric repulsion dominates and the polymer
coil takes a much more open structure, characterized by an
exponent $\nu \simeq 3/5$~\cite{Flory1}. The general picture that
emerges is that the typical spatial size of a polymer coil is much
smaller than the extended length $L=aN$ but larger than the size
of the ideal chain $aN^{1/2}$. The reason for this peculiar
behavior is entropy combined with the favorable interaction
between monomers and solvent molecules in good solvents. The
number of polymer configurations having a small end-to-end radius
is large, and these configurations are  entropically favored over
configurations characterized by a large end-to-end radius, for
which the number of possible polymer conformations is drastically
reduced. It is this conformational freedom of polymer coils which
leads to salient differences between polymer adsorption and that
of simple liquids.

Finally, in bad solvent conditions,
the polymer and the solvent are not compatible. A single
polymer chain collapses on itself in order to minimize
the monomer-solvent interaction. It is clear that
in this case, the polymer size, like any space filling object,
scales as $N \sim R_e^3$, yielding $\nu=1/3$.

\section{Single Chain Adsorption}
\setcounter{equation}{0}

Let us consider now the interaction of a single  polymer chain
with a solid substrate. The main effects particular to the
adsorption of polymers (as opposed to the adsorption of simple
molecules) are due to the reduction of conformational states of
the polymer at the substrate, which is due to the impenetrability
of the substrate for monomers~\cite{fse53}-\cite{jr77}.
The second factor determining the adsorption behavior is the
substrate-monomer interaction. Typically, for the case of an
adsorbing substrate, the interaction potential $V(z)$ between the
substrate and a single monomer has a form similar to the one shown
in Fig.~6, where $z$ measures the distance of the monomer from the
substrate surface,

\begin{equation}
\label{freepot}
V(z) \simeq
     \left\{ \begin{array}{llll}
     &  \infty
         & {\rm for} &  z<0  \\
     &  -U
         & {\rm for} &  0<z< B \\
     & -b z^{-\tau}
         & {\rm for }     & z>B \\
                \end{array} \right.
\end{equation}

\noindent The separation of $V(z)$ into three parts is done for
convenience. It consists of a hard wall at $z=0$, which embodies
the impenetrability of the substrate, \ie, $V(z) = \infty$ for
$z<0$. For positive $z$ we assume the potential to be given by an
attractive well of depth $U$ and width $B$. At large distances,
$z>B$, the potential can be modeled by a long-ranged attractive
tail decaying as $V(z) \sim -b z^{-\tau}$.

For the important case of (unscreened and non-retarded)
van-der-Waals interactions between the substrate and the polymer
monomers, the potential shows a decay governed by the exponent
$\tau=3$ and can be attractive or repulsive, depending on the
solvent, the chemical nature of the monomers and the substrate
material. The decay power $\tau=3$ follows from  the van-der-Waals
pair interaction, which decays as the inverse sixth power with
distance, by integrating over the three spatial dimensions of the
substrate, which is supposed to be a semi-infinite half
space~\cite{Israel}.

The strength of the potential well is measured by $U/(k_BT)$, \ie,
by comparing the potential depth $U$ with the thermal energy
$k_BT$. For strongly attractive  potentials, \ie, for $U$ large
or, equivalently, for low temperatures, the polymer is strongly
adsorbed and the thickness of the adsorbed layer, $D$,
approximately equals the potential range $B$. The resulting
polymer structure is shown in Fig.~7a, where the width of the
potential well, $B$, is denoted by a broken line.

For weakly attractive potentials, or for high temperatures, we anticipate a
weakly adsorbed polymer layer, with a diffuse layer thickness $D$ much
larger than the potential range $B$. This structure is depicted in Fig.~7b.
For both cases shown in Fig.~7, the polymer conformations are unperturbed
on a spatial scale of the order of $D$; on larger length scales, the
polymer is broken up into decorrelated {\em polymer
blobs}~\cite{degennes,Cloizeaux}, which are denoted by dashed circles in
Fig.~7. The idea of introducing polymer blobs is related to the fact that
very long and flexible chains have different spatial arrangement at small
and large length scales. Within each blob the short range interaction is
irrelevant, and the polymer structure inside the blob is similar to the
structure of an unperturbed polymer far from the surface. Since all
monomers are connected, the blobs themselves are linearly connected and
their spatial arrangement represents the behavior on large length scales.
In the adsorbed state, the formation of each blob leads to an entropy loss
of the order of one $k_BT$ (with a numerical prefactor of order unity which
is neglected in this scaling argument), so the total entropy loss of a
chain of $N$ monomers is ${\cal F}_{\rm rep} \sim k_BT (N/g)$, where $g$
denotes the number of monomers inside each blob.

Using the scaling relation $D \simeq a g^\nu$ for the blob size
dependence on the number of monomers $g$, the entropy penalty for
the confinement of a polymer chain to a width $D$ above the
surface can be written as~\cite{daoud77}

\be
\label{freerep}
\frac{{\cal F}_{\rm rep}}{k_BT} \simeq N \left(\frac{a}{D} \right)^{1/\nu}
\ee

\noindent The adsorption behavior of a  polymer chain results from
a competition between the attractive potential $V(z)$, which tries
to bind the monomers to the substrate, and the entropic repulsion
${\cal F}_{\rm rep}$, which tries to maximize entropy and,
therefore, favors a delocalized state where a large fraction of
the monomers are located farther away from the surface.

It is of interest to compare the adsorption of long--chain
polymers with the adsorption of small molecular solutes. Small
molecules adsorb onto a surface only if there is a bulk reservoir
with non-zero concentration in equilibrium with the surface. An
infinite polymer chain $N\rightarrow \infty$ behaves differently
as it remains adsorbed also in the limit of zero bulk
concentration. This corresponds to a true thermodynamic phase
transition in the limit
 $N\rightarrow \infty$ \cite{dg69}.
 For finite polymer length, however, the equilibrium
behavior is, in some sense, similar to the adsorption of small
molecules. A non-zero bulk polymer concentration will lead to
adsorption of polymer chains on the substrate. Indeed as all real
polymers are of finite length, the adsorption of single polymers
is never observed in practice. However, for fairly long polymers,
the desorption of a single polymer is almost a `true' phase
transition, and corrections due to finite (but long) polymer
length are often below experimental resolution.

\subsection{Mean--Field Regime}

Fluctuations of the local monomer concentration are of importance
to the description of polymers at surfaces due to the many
possible chain conformations. These fluctuations are treated
theoretically using field-theoretic or transfer-matrix techniques.
In a field-theoretic formalism, the problem of accounting for
different polymer conformations is converted into a functional
integral over different monomer-concentration
profiles~\cite{Cloizeaux}. Within transfer-matrix techniques, the
Markov-chain property of ideal polymers is exploited to re-express
the conformational polymer fluctuations as a product of
matrices~\cite{Flory2}.

However, there are cases where fluctuations in the local monomer
concentration become unimportant. Then, the adsorption behavior of
a single polymer chain is obtained using simple {\em mean--field
theory} arguments. Mean--field theory is a very useful
approximation applicable in many branches of physics, including
polymer physics. In a nutshell, each monomer is placed in a
``field", generated by the averaged interaction with all the other
monomers.

The mean--field theory can be justified for two cases: (i) a {\em
strongly adsorbed polymer chain}, \ie, a polymer chain which is
entirely confined inside the potential  well; and, (ii) the case
of {\em long-ranged attractive surface potentials}. To proceed, we
assume that the  adsorbed polymer layer is confined with an
average thickness $D$, as depicted in Fig.~7a or 7b. Within
mean--field theory, the polymer chain feels an average of the
surface potential, $\langle V(z) \rangle$, which is replaced by
the potential evaluated at the average distance from  the surface,
$\langle z \rangle \simeq D/2$. Therefore, $\langle V(z) \rangle
\simeq V(D/2)$. Further stringent conditions when such a
mean--field theory is valid are detailed below. The full free
energy of one chain, ${\cal F}$, of polymerization index $N$, can
be expressed as the sum of the repulsive entropic term,
Eq.~(\ref{freerep}), and the average potential

\be
\label{free1}
\frac{{\cal F}}{k_B T} \simeq N \left(\frac{a}{D} \right)^{1/\nu}
+N \frac{V(D/2)}{k_BT }
\ee

\noindent
Let us consider first the case of a strongly adsorbed polymer, confined
to a potential well of depth $~-U$. In this case the potential energy
per monomer becomes $V(D/2) \simeq -U$.
Comparing  the repulsive entropic term
with the potential term, we find
the two terms to be of equal strength for
a well depth $U^* \simeq k_BT (a/D)^{1/\nu}$.
Hence, the strongly adsorbed state, which is depicted in Fig.~7a,
should be realized for a high  attraction strength $U> U^* $.
For smaller attraction strength, $U < U^*$,
the adsorbed chain will actually be adsorbed
in a layer of width $D$ much larger
than the potential width $B$, as shown in Fig.~7b.
Since the  threshold energy $U^*$ is proportional to the temperature,
it follows that at high temperatures
it becomes increasingly difficult to confine
the chain. In fact, for an ideal chain,
with $\nu=1/2$, the resulting scaling relation for the critical well depth,
$U^* \sim k_BT (a/D)^2$, agrees with exact transfer-matrix predictions
for the adsorption threshold in  a square-well potential~\cite{square}.

We turn now to the case of a weakly adsorbed polymer layer. The
potential depth is smaller than the threshold, \ie, $U< U^*$, and
the stability of the weakly adsorbed polymer chains (depicted in
Fig.~7b) has to be examined. The thickness $D$ of this polymer
layer follows from the minimization of the free energy,
Eq.~(\ref{free1}),  with respect to $D$, where we use the
asymptotic form of the surface potential, Eq.~(\ref{freepot}), for
large separations.  The result is
\be
\label{Dmean}
D \simeq \left( \frac{a^{1/\nu} k_BT}{b} \right)^{\nu/(1-\nu \tau)}
\ee
Under which circumstances is the prediction Eq. (\ref{Dmean})
correct, at least on a qualitative level? It turns out that the
prediction for $D$, Eq.~(\ref{Dmean}), obtained within the simple
mean--field theory, is correct if the attractive tail of the
substrate potential in Eq.~(\ref{freepot}) decays for large values
of $z$ slower than the entropic repulsion in Eq.~(\ref{freerep})
~\cite{lip89}. In other words, the mean--field theory is valid for
weakly-adsorbed polymers only for $\tau < 1/\nu$. This can already
be guessed from the functional form of  the layer thickness, Eq.
(\ref{Dmean}), because for $\tau > 1/ \nu$ the layer thickness $D$
goes to zero as $b$ diminishes. Clearly an unphysical result. For
ideal polymers (theta solvent, $\nu=1/2$), the validity condition
is $\tau<2$, whereas for swollen polymers (good solvent
conditions, $\nu = 3/5$), it is $\tau < 5/3$. For most
interactions (including van der Waals interactions with $\tau =
3$) this condition on $\tau$ is not satisfied, and fluctuations
are in fact important, as is discussed in the next section.

There are two notable exceptions. The first is for charged
polymers close to an oppositely charged surface, in the {\it
absence} of salt ions. Since the attraction of the polymer to an
infinite, planar and charged surface is  linear in $z$, the
interaction is described by Eq.~(\ref{freepot}) with  an exponent
$\tau=-1$, and the inequality $\tau<1/\nu$ is satisfied. For
charged surfaces, Eq.~(\ref{Dmean}) predicts the thickness $D$ to
increase to infinity as the temperature increases or as the
attraction strength $b$ (proportional to the surface charge
density) decreases. The resultant exponents for the scaling of $D$
follow from Eq. (\ref{Dmean}) and are $D\sim (T/b)^{1/3}$ for
ideal chains, and $D\sim (T/b)^{3/8}$ for swollen
chains~\cite{netz95}-\cite{xav98}.

A second example where the mean--field theory can be used is the
adsorption of polyampholytes on charged surfaces~\cite{polyam1}.
Polyampholytes are polymers consisting of negatively and
positively charged monomers. In cases where the total charge on
such a polymer adds up to zero, it might seem that the interaction
with a charged surface should vanish. However, it turns out that
local charge fluctuations (\ie, local spontaneous dipole moments)
lead to a strong attraction of polyampholytes to charged
substrates. In  absence of salt this attractive interaction has an
algebraic decay with an exponent $\tau = 2$ \cite{polyam1}. On the
other hand, in the presence of salt, the effective interaction is
exponentially screened, yielding a decay faster than the
fluctuation repulsion, Eq.~(\ref{freerep}). Nevertheless, the
mean--field theory, embodied in the free energy expression
Eq.~(\ref{free1}), can be used to predict the adsorption phase
behavior within the strongly adsorbed case (\ie, far from any
desorption transition)~\cite{polyam2}.

\subsection{Fluctuation Dominated Regime}

Here we consider  the weakly adsorbed case for substrate
potentials which decay (for large separations from the surface)
faster than the entropic repulsion Eq.~(\ref{freerep}), \ie, $\tau
> 1/\nu$. This applies, \eg, to van-der-Waals attractive
interaction between the substrate and monomers, screened
electrostatic interactions,
 or any other short-ranged potential. In this case,
fluctuations play a decisive role. In fact, for {\em ideal
chains}, it can be rigorously proven (using transfer-matrix
techniques) that all potentials decaying faster than $z^{-2}$ for
large $z$ have a  continuous adsorption transition at a finite
critical temperature $T^*$ \cite{lip89}. This means that the
thickness of the adsorbed polymer layer diverges for $T
\rightarrow T^*$ as

\be
\label{Dideal}
D \sim (T^*-T)^{-1}
\ee

\noindent The power law divergence of $D$ is universal. Namely, it
does not depend on the specific functional form and strength of
the potential as long as they satisfy the above condition.

The case of {\em non-ideal chains} is much more complicated. First
progress has been made by de Gennes who recognized the analogy
between the partition function of a self-avoiding chain and the
correlation function of an $n$-component spin model in the
zero-component ($n\rightarrow 0$) limit \cite{gennes72}. The
adsorption behavior of non-ideal chains has been treated by
field-theoretic methods  using the analogy to surface critical
behavior of magnets (again in the $n\rightarrow 0$ limit)
\cite{erich,ekb82}. The resulting behavior is similar to the
ideal-chain case and shows an adsorption transition at a finite
temperature, and a continuous increase towards infinite layer
thickness characterized by a power law divergence as function of
$T-T^*$ \cite{ekb82}.

The complete behavior for ideal and swollen chains can be
described using scaling ideas in the following way. The entropic
loss due to the confinement of the chain to a region of thickness
$D$ close to the surface is  again given by Eq.~~(\ref{freerep}).
Assuming that the adsorption layer is much thicker than  the range
of the attractive potential  $V(z)$, the attractive potential can
be assumed to be localized  at the substrate surface $V(z)\simeq
V(0)$. The attractive free energy of the chain to the substrate
surface can then be written as~\cite{dg76}

\be
\label{freeatt1} {\cal F}_{\rm att}\simeq - \tilde{\gamma}k_B
(T^*-T) N f_1 = - \gamma_1 a^2 N f_1 \ee

\noindent where $f_1$ is the probability to find a monomer at the
substrate surface and $\tilde{\gamma}$ is a dimensionless
interaction parameter. Two surface excess energies are typically
being used: $\gamma_1=\tilde{\gamma}k_B(T^*-T)/ a^2$ is the excess
energy per unit area, while  $\gamma_1 a^2$ is the excess energy
per monomer at the surface. Both are positive for the attractive
case (adsorption) and negative for the depletion case. The
dependence of $\gamma_1$ on $T$ in Eq.~~(\ref{freeatt1}) causes
the attraction to vanish at a critical temperature, $T=T^*$, in
accord with our expectations.

The contact probability for a swollen chain with the surface,
$f_1$, can be calculated as follows~\cite{dgp83}. In order to force
the chain of polymerization index $N$ 
to be in contact with the wall, one of the chain ends is pinned to the
substrate.
The number of monomers which are in contact with the surface can
be calculated using field-theoretic methods and is given by
$N^{\varphi}$, where $\varphi$ is called the {\em surface
crossover exponent}~\cite{erich,ekb82}. The fraction of bound
monomers follows to be $f_1 \sim N^{\varphi-1} $, and thus goes to
zero as the polymer length increases, for $\varphi<1$. Now instead
of speaking of the entire chain, we refer to a `chain of blobs'
(See Fig.~7) adsorbing on the surface, each blob consisting of $g$ monomers. 
We proceed by assuming that
the size of an adsorbed blob $D$ scales with the number of monomers
per blob $g$ similarly as in the bulk, $D \sim a g^{\nu}$, as is
indeed confirmed by field theoretic calculations. The fraction of
bound monomers can be expressed in terms of $D$ and is given by

\be
\label{fprob1}
f_1 \sim \left(\frac{D}{a}\right)^{(\varphi-1)/\nu}
\ee

\noindent
Combining the entropic repulsion, Eq.~(\ref{freerep}),
and the substrate attraction, Eqs.~(\ref{freeatt1}-\ref{fprob1}),
the total free energy is given by

\be
\label{free2} \frac{{\cal F}}{k_BT} \simeq N \left(\frac{a}{D}
\right)^{1/\nu}- N\frac{\tilde{\gamma}(T^* - T)}{T}
\left(\frac{D}{a}\right)^{(\varphi-1)/\nu} \ee

\noindent
Minimization with respect to $D$ leads to the final result

\be
\label{Dnonideal} D \simeq a \left[\frac{\tilde{\gamma}
(T^*-T)}{T}\right]^{-\nu / \varphi} \simeq a \left(\frac{\gamma_1
a^2}{k_B T}\right)^{-\nu / \varphi} \ee

\noindent
For ideal chains, one has $\varphi=\nu=1/2$, and thus we recover
the prediction from the transfer-matrix calculations,
Eq.~(\ref{Dideal}).
For non-ideal chains, the crossover
exponent $\varphi$ is in general different from the swelling
exponent $\nu$. However, extensive Monte Carlo
computer simulations point to a value for $\varphi$ very close to
$\nu$,  such that the adsorption exponent $\nu/\varphi$
appearing in Eq.~(\ref{Dnonideal}) is very close to unity,
for polymers embedded in three dimensional space~\cite{ekb82}.

A further point which has been calculated using field theory is
the behavior of the monomer volume fraction $\phi(z)$ close to the
substrate. From rather general arguments borrowed from the theory
of critical phenomena, one expects a power-law behavior for
$\phi(z)$ at sufficiently small distances from the substrate
~\cite{ekb82,dgp83,bd87}

\be
\label{proxscale} \phi(z) \simeq  \phi_s (z/a)^m
\ee
\noindent
recalling that the monomer density is related to $\phi(z)$ by
$c(z)=\phi(z)/a^3$.

In the following, we  relate the so-called {\em proximal exponent}
$m$ with the two other exponents introduced above, $\nu$ and
$\varphi$. First note that the surface value of the monomer volume
fraction, $\phi_s = \phi(z \approx a)$, for one adsorbed blob
follows from the number of monomers at the surface per blob, which
is given by $f_1 g$, and the cross-section area of a blob, which
is of the order of $D^2$. The surface volume fraction is given by

\be
\label{surf}
\phi_s \sim \frac{ f_1 g a^2}{D^2} \sim g^{\varphi-2 \nu}
\ee

\noindent
Using the scaling prediction Eq.~(\ref{proxscale}), we see that
the monomer volume fraction at  the blob center, $z \simeq D/2$,
is given by $\phi(D/2) \sim  g^{\varphi-2 \nu} (D/a)^m$,
which (again using $D  \sim a g^\nu$) can be rewritten as
$\phi(D/2) \simeq g^{\varphi-2 \nu+m\nu}$.

On the other hand, at a distance $D/2$ from the surface, the
monomer volume fraction should have decayed to the average monomer
volume fraction $a^3 g/D^3 \sim  g^{1-3\nu}$ inside the blob since
the statistics of the chain inside the blob is like for a chain in the bulk.
By direct comparison of the two volume fractions, we see that the
exponents $\varphi-2 \nu+m\nu$ and $1-3\nu$ have to match in order
to have a consistent result, yielding

\be
m = \frac{1-\varphi-\nu}{\nu}
\label{proxm}
\ee

\noindent
 For ideal chain (theta solvents), one has
$\varphi=\nu=1/2$. Hence, the proximal exponent vanishes, $m=0$.
This means that the proximal exponent has no mean--field analog,
explaining  why it was discovered only  within field-theoretic
calculations~\cite{erich,ekb82}. In the presence of correlations
(good solvent conditions) one has $ \varphi \simeq \nu \simeq 3/5$
and thus $m \simeq 1/3$.

Using $D \simeq a g^\nu$ and Eq.~(\ref{Dnonideal}),
the surface volume fraction, Eq.~(\ref{surf}),
can be rewritten as

\be
\phi_s \sim \left( \frac{D}{a} \right)^{(\varphi-2\nu)/\nu} \sim
\left( \frac{\gamma_1}{k_BT} \right)^{(2\nu-\varphi)/\varphi}
\simeq \frac{\gamma_1}{k_BT} \label{phi213}
\ee

\noindent where in the last approximation appearing in
Eq.~(\ref{phi213}) we used the fact that $\varphi \simeq \nu$. The
last result shows that the surface volume fraction within one blob
can become large if the adsorption energy $\gamma_1$ is large
enough as compared with $k_B T$. Experimentally, this is very
often the case, and additional interactions (such as multi-body
interactions) between monomers at
the surface have in principle to be taken into account.

After having discussed the adsorption behavior of a single chain,
a word of caution is in order. Experimentally, one never looks at
single chains adsorbed to a surface. First, this is due to the
fact that one always works with polymer solutions, where there is
a large number of polymer chains contained in the bulk reservoir,
even when the bulk monomer (or polymer) concentration is quite
low. Second, even if the bulk polymer concentration is very low,
and in fact so low that polymers in solution rarely interact with
each other, the surface concentration of polymer is enhanced
relative to that in the bulk. Therefore, adsorbed polymers at the
surface usually do interact with neighboring chains, due to the
higher polymer  concentration at the surface~\cite{bd87}.

Nevertheless, the adsorption behavior of a single chain serves as
a basis and guideline for the more complicated adsorption
scenarios involving many-chain effects. It will turn out that the
scaling of the adsorption layer thickness $D$ and the proximal
volume fraction profile, Eqs.~(\ref{Dnonideal}) and
(\ref{proxscale}), are not affected by the presence of other
chains. This finding as well as other many-chain effects on
polymer adsorption is the subject of the next section.


\section{Polymer Adsorption from Solution}
\setcounter{equation}{0}


\subsection{The Mean--Field Approach: Ground State Dominance}

In this section we look at the equilibrium behavior of many chains
adsorbing on (or equivalently depleting from) a surface in contact
with a bulk reservoir of chains at equilibrium. The polymer chains
in the reservoir are assumed to be in a semi-dilute concentration
regime. The semi-dilute regime is defined by  $c> c^{*}$, where
$c$ denotes the monomer concentration (per unit volume) and
$c^{*}$ is the concentration where individual chains start to
overlap. Clearly, the overlap concentration is reached when the
average bulk monomer concentration exceeds the monomer
concentration inside a polymer coil. To estimate the overlap
concentration $c^*$, we simply note that the average monomer
concentration inside a coil with dimension $R_e \sim a N^\nu$ is
given by $c^* \sim N/ R_e^3 \sim N^{1-3\nu} /a^3$.

As in the previous section, the adsorbing surface is taken as an
ideal and smooth plane. Neglecting lateral concentration
fluctuations, one can reduce the problem to an effective
one-dimensional problem, where the monomer concentration depends
only on the distance $z$ from the surface, $c=c(z)$. The two
boundary values are: $c_b=c(z\rightarrow \infty)$ in the bulk,
while  $c_s=c(z=0)$ on the surface.

In addition to the monomer concentration $c$, it is more
convenient to work with the monomer volume fraction: $\phi(z)=a^3
c(z)$ where $a$ is the monomer size. While the bulk value (far
away from the surface) is fixed by the concentration in the
reservoir, the value on the surface at $z=0$ is self-adjusting in
response  to a given surface interaction. The simplest
phenomenological surface interaction is linear  in the surface
polymer concentration. The resulting contribution to the surface
free energy (per unit area) is

\be
F_s=-\gamma_1\phi_s
\label{d1}
\ee

\noindent where $\phi_s=a^3 c_s$ and a positive (negative)  value
of $\gamma_1=\tilde{\gamma}k_B (T-T^*)/a^2$,
 defined in the previous section,
enhances adsorption (depletion) of the chains on (from) the
surface. However, $F_s$ represents only the local reduction in the
interfacial free energy due to the adsorption. In order to
calculate the full interfacial free energy, it is important to
note that monomers adsorbing on the surface are connected to other
monomers belonging to the same polymer chain. The latter
accumulate in the vicinity of the surface. Hence, the interfacial
free energy does not only depend on the surface concentration of
the monomers but also on their concentration in the {\it vicinity}
of the surface. Due to the polymer flexibility and connectivity,
the entire adsorbing layer can have a considerable width. The {\it
total} interfacial free energy of the polymer chains will depend
on this width and is quite different from the interfacial free
energy for simple molecular liquids.

There are several theoretical approaches
to treat this polymer adsorption. One of the simplest approaches
which yet gives reasonable qualitative results is
the Cahn -- deGennes approach~\cite{ch58,dg81}.
In this approach, it is possible to write down a continuum functional
which describes the contribution to the free energy of the
polymer chains in the solution. This procedure was introduced
by Edwards in the 60's ~\cite{e65}
and was applied to polymers at interfaces
by de Gennes~\cite{dg81}.
Below we present such a continuum version which can be studied
analytically. Another approach is a discrete one, where the monomers
and solvent molecules are put on a lattice. The latter approach
is quite useful in computer simulations and numerical self consistent field
(SCF) studies and is reviewed elsewhere~\cite{fleer}.

In the continuum approach and using a mean--field theory, the bulk
contribution to the adsorption free energy is written in terms of
the local monomer volume fraction $\phi(z)$, neglecting all kinds
of monomer-monomer correlations. The total reduction in the
surface tension (interfacial free energy per unit area) is then

\bea
\gamma-\gamma_0 = -\gamma_1\phi_s + \int_{0}^{\infty}
\dd z \Bigl[ L(\phi)\Bigl({d\phi\over{dz}}\Bigr)^2 + F(\phi)-F(\phi_b)
+\mu(\phi-\phi_b) \Bigr]
\label{d2}
\eea

\noindent where $\gamma_0$ is the bare surface tension of the
surface in contact with the solvent but without the presence of
the monomers in solution, and $\gamma_1$ was defined in
Eq.~(\ref{d1}). The stiffness function $L(\phi)$ represents the
energy cost of local concentration fluctuations and its form is
specific to long polymer chains. For low polymer concentration it
can be written as~\cite{degennes}:

\bea
L(\phi)={k_B T\over a^3}\Bigl({a^2\over 24\phi}\Bigr)
\label{d3}
\eea

\noindent
where $k_B T$ is the thermal energy. The other terms in Eq.~(\ref{d2})
come from the Cahn-Hilliard free energy of mixing of the
polymer solution, $\mu$ being the chemical potential, and~\cite{Flory1}

\bea
F(\phi)={k_B T\over a^3}\Bigl( {\phi\over N}\log \phi + \half v\phi^2
+{1\over 6} w\phi^3+ \cdots \Bigr)
\label{d4}
\eea

\noindent where $N$ is the polymerization index. In the following,
we neglect the first term in Eq.~(\ref{d4}) (translational
entropy), as can be justified in the long chain limit, $N\gg 1$.
The second and third dimensionless virial coefficients are $v$ and
$w$, respectively. Good, bad and theta solvent conditions are
achieved, respectively, for positive, negative or zero $v$. We
concentrate hereafter only on good solvent conditions, $v>0$, in
which case the higher order $w$-term can be safely neglected. In
addition, the local monomer density is assumed to be small enough,
in order to justify the omission of higher virial coefficients.
Note that for small molecules the translational entropy always
acts in favor of desorbing from the surface. As was discussed in
the Sect. 1, the vanishing small translational entropy for
polymers results in a stronger adsorption (as compared with small
solutes) and makes the polymer adsorption much more of an
irreversible process.

The key feature in obtaining Eq.~(\ref{d2}) is the so-called {\em
ground state dominance}, where for long enough chains $N\gg 1$,
only the lowest energy eigenstate (ground state) of a
diffusion-like equation is taken into account. This approximation
gives us the leading behavior in the $N\rightarrow \infty$
limit~\cite{dg69}. It is based on the fact that the weight of the
first excited eigenstate is smaller than that of the ground state
by an exponential factor: $\exp(-N\,\Delta E)$ where $\Delta
E=E_1-E_0>0$ is the difference in the eigenvalues between the two
eigenstates. Clearly, close to the surface more details on the
polymer conformations can be important. The adsorbing chains have
tails (end-sections of the chains that are connected to the
surface by only one end), loops (mid-sections of the chains that
are connected to the surface by both ends), and trains (sections
of the chains that are adsorbed on the surface), as depicted in
Fig.~3a. To some extent it is possible to get profiles of the
various chain segments even within mean--field theory, if the
ground state dominance condition is relaxed as is discussed below.

Taking into account all those simplifying assumptions and
conditions, the mean--field theory for the interfacial free energy
can be written as:

\bea
\gamma-\gamma_0 = -\gamma_1\phi_s+{k_B T\over a^3}
\int_{0}^{\infty}\dd z\,\,\Bigl[{a^2\over 24\phi}
{\Bigl({\dd\phi\over \dd z}\Bigr)}^2
+ \half v (\phi(z)-\phi_b)^2\Bigr]
\label{d5}
\eea

\noindent where the monomer bulk chemical potential $\mu$ is given
by $\mu=\partial F/\partial \phi|_b =v\phi_b$.

It is also useful to define
the total amount of monomers per unit
area which take part in the adsorption layer. This is the so-called
surface excess $\Gamma$; it is measured experimentally
using, \eg, ellipsometry, and is defined as

\bea
\Gamma={1\over a^3}\int_0^\infty \dd z\,\,[\phi(z)-\phi_b]
\label{d6}
\eea

\noindent
The next step is to minimize the free energy functional
(\ref{d5}) with respect to both $\phi(z)$ and $\phi_s=\phi(0)$.
It is more convenient to re-express Eq.~(\ref{d5}) in terms of
$\psi(z)=\phi^{1/2}(z)$
and $\psi_s=\phi_s^{1/2}$

\bea
\gamma-\gamma_0 = -\gamma_1\psi_s^2+{k_B T\over a^3}
\int_{0}^{\infty}\dd z\,\,\Bigl[{a^2\over 6}
{\Bigl({\dd\psi\over \dd z}\Bigr)}^2
+ \half v (\psi^2(z)-\psi_b^2)^2\Bigr]
\label{d6a}
\eea

\noindent Minimization of Eq.~(\ref{d6a}) with respect to
$\psi(z)$ and $\psi_s$ leads to the following profile equation and
boundary condition

\bea
{a^2\over 6}{\dd^2\psi\over \dd z^2} &=& v\psi(\psi^2-\psi_b^2)
\nonumber\\
& & \nonumber \\
{{1\over \psi_s}{\dd \psi\over \dd z}}\Bigl|_s &=&
-{6a\over k_B T}\gamma_1=-{1\over 2D}
\label{d7}
\eea

\noindent The second equation sets a boundary condition on the
logarithmic derivative of the monomer volume fraction, $\dd
\log\phi/\dd z|_s =2\psi^{-1}\dd \psi/\dd z|_s=-1/D$, where the
strength of the surface interaction $\gamma_1$ can be expressed in
terms of a length $D\equiv k_B T/(12 a\gamma_1)$. Note that
exactly the same scaling of $D$ on $\gamma_1/T$ is obtained  in
Eq.~(\ref{Dnonideal}) for the single chain behavior if one sets
$\nu = \varphi = 1/2$ (ideal chain exponents). This  is strictly
valid at the upper critical dimension ($d=4$) and is a very good
approximation in three dimensions.

The profile equation (\ref{d7}) can  be integrated
once, yielding

\be
{a^2\over 6}\left({\dd\psi\over \dd z}\right)^2
= \half v(\psi^2-\psi_b^2)^2
\label{d7a}
\ee

\noindent
The above differential equation can now be solved analytically for
adsorption ($\gamma_1>0$) and depletion ($\gamma_1<0$).

We first present the results in more detail for polymer adsorption
and then repeat the main findings for polymer depletion.

\subsubsection{Polymer Adsorption}

Setting $\gamma_1>0$ as is applicable for the adsorption case, the
first-order differential equation (\ref{d7a}) can be integrated
and together with the boundary condition Eq.~(\ref{d7}) yields

\bea
\phi(z)&=&\phi_b\coth^2\Bigl({z+z_0\over\xi_b}\Bigr)
\label{d8ad}
\eea

\noindent where the length $\xi_b=a/\sqrt{3v \phi_b}$ is the
Edwards correlation length characterizing the exponential decay of
concentration fluctuations in the bulk~\cite{degennes,e65}. The
length $z_0$ is not an independent length since it depends on $D$
and $\xi_b$, as can be seen from the boundary condition
Eq.~(\ref{d7})

\bea
z_0={\xi_b\over 2}{\rm arcsinh} \Bigl({4D\over \xi_b}\Bigr)=
\xi_b {\rm arccoth} (\sqrt{\phi_s/\phi_b})
\label{d9}
\eea

\noindent
Furthermore, $\phi_s$ can be directly related to the
surface interaction $\gamma_1$ and the bulk value $\phi_b$

\be
\frac{\xi_b}{2D}=\frac{6a^2\gamma_1}{k_B T \sqrt{3v\phi_b}}=
\sqrt{\frac{\phi_b}{\phi_s}}\left(\frac{\phi_s}{\phi_b}-1\right)
\label{d10}
\ee

In order to be consistent with the semi-dilute concentration
regime, the correlation length $\xi_b$ should be smaller than the
size of a single chain, $R_e=a N^{\nu}$, where $\nu= 3/5$ is the
Flory exponent in good solvent conditions. This sets a lower bound
on the polymer concentration in the bulk, $c>c^*$.

So far three length scales have been introduced: the Kuhn length
or monomer size $a$, the adsorbed-layer width $D$, and the bulk
correlation length  $\xi_b$. It is more convenient for the
discussion to consider the case where those three length scales
are quite separated: $a\ll D \ll \xi_b$. Two conditions must be
satisfied. On one hand, the adsorption parameter is not large, $12
a^2 \gamma_1\ll k_B T$ in order to have $D\gg a$. On the other
hand, the adsorption energy is large enough to satisfy $12 a^2
\gamma_1\gg k_B T \sqrt{3 v \phi_b}$ in order to have $D\ll
\xi_b$. The latter inequality can be regarded also as a condition
for the polymer bulk concentration. The bulk correlation length is
large enough if indeed the bulk concentration (assumed to be in
the semi-dilute concentration range) is not too large. Roughly,
let us  assume in a typical case that the three length scales are
well separated: $a$ is of the order of a few Angstroms, $D$ of the
order of a few dozens of Angstroms, and $\xi_b$ of the order of a
few hundred Angstroms.

When the above two inequalities are satisfied,
three spatial regions of adsorption can be differentiated:
the proximal, central, and distal regions,
 as is outlined below. In addition, as soon as $\xi_b \gg D$,
$z_0\simeq 2 D$, as follows  from Eq.~(\ref{d9}).

\begin{itemize}
\item
Close enough to the surface, $z \sim a$, the adsorption profile
depends on the details of the short range interactions between the
surface and monomers. Hence, this region is not universal. In the
proximal region, for $a\gg z\gg D$, corrections to the mean--field
theory analysis  (which assumes the concentration to be constant)
are presented below similarly to the treatment of the single chain
section. These corrections reveal a new scaling exponent
characterizing the concentration profile. They are of particular
importance close to the adsorption/desorption transition.

\item
In the distal region, $z\gg\xi_b$, the excess polymer
concentration decays exponentially to its bulk value

\be
\phi(z)-\phi_b \simeq 4 \phi_b \e^{-2z/\xi_b}
\label{d11}
\ee

\noindent as follows from Eq.~(\ref{d8ad}). This behavior is very
similar to the decay of fluctuations in the bulk with $\xi_b$
being the correlation length.

\item
Finally, in the central region (and with the assumption that
$\xi_b$ is the largest length scale in the problem), $ D \ll z \ll
\xi_b$, the profile is universal and from Eq.~(\ref{d8ad}) it can
be shown to decay with a power law

\bea
\phi(z)&=& {1\over 3v}{\Bigl({a\over z+2D}\Bigr)}^2
\label{d12ad}
\eea

\end{itemize}

\noindent A sketch of the different scaling regions in the
adsorption profile is given in Fig.~8a. Included in this figure
are corrections in the proximal region, which is discussed further
below.

A special consideration should be given to the formal limit of
setting the bulk concentration to zero, $\phi_b \rightarrow 0$
(and equivalently $\xi_b \rightarrow \infty$), which denotes the limit
of an adsorbing layer in contact with a polymer
reservoir of vanishing concentration.
It should be emphasized that this limit  is not
consistent with the assumption of a semi-dilute polymer solution
in the bulk. Still,  some information on the polymer density
profile close to the adsorbing surface, where the polymer solution
is locally semi-dilute~\cite{bd87} can be obtained.
Formally, we
take the limit $\xi_b \rightarrow \infty$ in Eq.~(\ref{d8ad}), and
the limiting expression given by Eq.~(\ref{d12ad}), does not
depend on $\xi_b$. The profile in the central region decays
algebraically. In the case of zero polymer concentration in the
bulk, the natural cutoff is not $\xi_b$ but rather $R_e$, the coil
size of a single  polymer in solution. Hence, the distal region
looses its meaning and is replaced by a more complicated scaling
regime~\cite{g92}. The length $D$ can be regarded as the layer
thickness in the $\xi_b\rightarrow \infty$ limit in the sense that
a finite fraction of all the monomers are located in this layer of
thickness $D$ from the surface. Another observation is  that
$\phi(z) \sim 1/z^2$ for $z\gg D$. This power law is a result of
the mean--field theory  and its modification is discussed below.

It is now possible to calculate within the mean--field theory the
two physical quantities that are measured in many experiments: the
surface tension reduction $\gamma-\gamma_0$ and the surface excess
$\Gamma$.

The surface excess, defined in Eq.~(\ref{d6}), can be calculated
in a close form by inserting Eq.~(\ref{d8ad}) into Eq.~(\ref{d6}),

\bea
\Gamma={1\over \sqrt{3v}a^2}\Bigl(\phi_s^{1/2}-\phi_b^{1/2}\Bigr)
={\xi_b\phi_b\over a^3}\Bigl(\sqrt{\phi_s\over \phi_b}-1\Bigr)
\label{d13}
\eea

\noindent For strong adsorption, we obtain from Eq.~(\ref{d10})
that  $\phi_s \simeq (a/2D)^2/3 v\gg \phi_b$, and Eq.~(\ref{d13})
reduces to

\bea
\Gamma ={1\over 3v a^2}\Bigl({a\over D}\Bigr) \sim \gamma_1
\label{d14}
\eea

\noindent while the surface volume fraction scales as $\phi_s\sim
\gamma_1^2$. As can be seen from Eqs.~(\ref{d14}) and
(\ref{d12ad}), the surface excess as well as the entire profile
does not depend (to leading order) on the bulk concentration
$\phi_b$. We note again that the strong adsorption condition is
always satisfied in the $\phi_b \rightarrow 0$ limit. Hence,
Eq.~(\ref{d14}) can be obtained directly by integrating the
profile in the central region, Eq. (\ref{d12ad}).

Finally, let us calculate the reduction  in surface tension
for the adsorbing case.
Inserting the variational
equations (\ref{d7}) in Eq.~(\ref{d5})
yields

\bea
\gamma-\gamma_0=- \gamma_1\phi_s+{k_B T \sqrt{3v}\over 9a^2}\phi_s^{3/2}
\Bigl[1-3\Bigl({\phi_b\over\phi_s}\Bigr)+
2\Bigl({\phi_b\over\phi_s}\Bigr)^{3/2}\Bigr]
\label{d16}
\eea

\noindent
The surface term in Eq.~(\ref{d16}) is negative
while the second term is positive.
For strong adsorption this reduction of $\gamma$
does not depend on $\phi_b$
and reduces to

\bea \gamma-\gamma_0\sim -{\left({\gamma_1 a^2\over k_B
T}\right)}^3~ {k_B T\over a^2} + {\cal O}(\gamma_1^{4/3})
\label{d17} \eea

\noindent
where the leading term is just the contribution of
the surface monomers.

\subsubsection{Polymer Depletion}
We highlight the main differences between  the polymer adsorption
and polymer depletion. Keeping in mind that $\gamma_1<0$ for
depletion, the solution of the same profile equation (\ref{d7a}),
with the appropriate boundary condition results in

\bea
\phi(z)&=&\phi_b\tanh^2\Bigl({z+z_0\over\xi_b}\Bigr)
\label{d8de}
\eea
%
which is schematically plotted in Fig.~8b.
The limit $\phi_b \rightarrow 0$ cannot be taken
in the depletion case since
depletion with respect to a null reservoir has no meaning.
However, we can, alternatively, look at the strong depletion limit,
defined by the condition $\phi_s \ll \phi_b$. Here
we find
\bea
\phi(z)&=& {3v\phi_b^2}{\Bigl({ z+2D\over a}\Bigr)}^2
\label{d12de}
\eea
%
In the same limit, we find for the surface volume fraction $\phi_s
\sim \phi_b^2\gamma_1^{-2}$, and the exact expression for the
surface excess Eq.~(\ref{d13}) reduces to

\bea
\Gamma  =-{1\over a^2}\sqrt{\phi_b\over 3v}\simeq -{\phi_b\xi_b\over a^3}
\label{d15}
\eea

\noindent
The negative surface excess  can be
directly estimated from a profile
varying from $\phi_b$ to zero over a length scale of order $\xi_b$.

The dominating behavior for the surface tension can be calculated
from Eq.~(\ref{d5}) where both terms are now positive. For the
strong depletion case we get

\bea
\gamma-\gamma_0 \simeq {k_B T \over a^2}\Bigl({a\over \xi_b}\Bigr)^3
\sim \phi_b^{3/2}
\label{d18}
\eea


\subsection{Beyond Mean--Field Theory: Scaling Arguments for Good Solvents}

One of the mean--field theory results that should be corrected is
the scaling of the correlation length with $\phi_b$. In the
semi-dilute regime, the correlation length can be regarded as the
average mesh size created by the overlapping chains. It can be
estimated using very simple scaling arguments~\cite{degennes}: The
volume fraction of monomers inside a coil formed by a subchain
consisting of $g$  monomers is $\phi \sim g^{1-3\nu}$ where $\nu$
is the Flory exponent. The spatial scale of this subchain is given
by $\xi_b \sim a g^\nu$. Combining these two relations, and
setting $\nu = 3/5$, as appropriate for good solvent conditions,
 we obtain the known scaling of
the correlation length

\be
\xi_b \simeq a\phi_b^{-3/4}
\label{d19}
\ee

\noindent This relation corrects the  mean--field theory result
$\xi_b \sim \phi_b^{-1/2}$  which can be obtained from, \eg,
Eq.~(\ref{d5}).

\subsubsection{Scaling for Polymer Adsorption}

We repeat here an argument due to de Gennes~\cite{dg81}. The main
idea is to assume that the relation Eq.~(\ref{d19}) holds locally:
$\phi(z)=[\xi(z)/a]^{-4/3}$, where $\xi(z)$ is the local ``mesh
size'' of the semi-dilute polymer solution. Since there is no
other length scale in the problem beside the distance from the
surface, $z$, the correlation length $\xi(z)$ should scale as the
distance $z$ itself, $\xi(z)\simeq z$ leading to the profile

\be
\phi(z) \simeq \left(\frac{a}{z}\right)^{4/3}
\label{d22}
\ee

\noindent We note that this argument holds only in the central
region $D\ll z\ll \xi_b$. It has been confirmed experimentally
using neutron scattering~\cite{neutscatt} and neutron
reflectivity~\cite{neutrefl}. Equation (\ref{d22}) satisfies the
distal boundary condition: $z\rightarrow \xi_b$, $\phi(z)
\rightarrow \phi_b$, but for $z > \xi_b$ we expect the regular
exponential decay behavior of the distal region, Eq.~(\ref{d11}).
De Gennes also proposed (without a rigorous proof) a convenient
expression for $\phi(z)$, which has the correct crossover from the
central to the mean--field proximal region~\cite{dg81}

\be
\phi(z)=\phi_s\left(\frac{{4\over3}D}{z+{4\over3}D}\right)^{4/3}
\simeq \left(\frac{a}{z+{4\over 3}D}\right)^{4/3}
\label{d23}
\ee

\noindent
Note that the above equation reduces to Eq.~(\ref{d22})
for $z\gg D$. The extrapolation of Eq.~(\ref{d23}) also gives the
correct definition of $D$: $D^{-1}= -\dd \log \phi/\dd z|_s$. In
addition, $\phi_s$ is obtained from the extrapolation to $z=0$ and
scales as

\be
\phi_s= \phi(z=0)=\left(\frac{a}{D}\right)^{4/3}
\label{d24}
\ee

\noindent
For strong adsorption ($\phi_s\gg \phi_b$), we have

\bea
\phi_s &\simeq& {\left(\frac{a}{D}\right)}^{4/3} \sim \gamma_1^2 \nonumber \\
&&\nonumber\\
D &\simeq& a \left(\frac{k_B T}{a^2\gamma_1}\right)^{3/2} \sim \gamma_1^{-3/2}
\nonumber\\
&&\nonumber\\
\Gamma &\simeq & a^2\left(\frac{a^2\gamma_1}{k_B T}\right)^{1/2} \sim
\gamma_1^{1/2}\nonumber \\
&&\nonumber \\
\gamma-\gamma_0 &\simeq &-\frac{k_B T}{a^2}\phi_s^{3/2} \sim -\gamma_1^3
\label{d25}
\eea

\noindent It is interesting to note that although $D$ and $\Gamma$
have different scaling with the surface interaction $\gamma_1$ in
the mean--field theory and scaling approaches, $\phi_s$ and
$\gamma-\gamma_0$ have the same scaling using both approaches.
This is a result of the same scaling $\phi_s \sim \gamma_1^2$,
which, in turn, leads to $\gamma-\gamma_0\simeq \gamma_1\phi_s\sim
\gamma_1^3$.

\subsubsection{Scaling for Polymer Depletion}

For polymer depletion similar arguments led de Gennes~\cite{dg81} to propose
the following scaling form for the central and mean--field proximal
regions, $a<z<\xi_b$,

\bea
\phi(z)=\phi_b\left( \frac{z+{5\over 3}D}{\xi_b} \right)^{5/3}
\label{d26}
\eea
%
where the depletion thickness is $\xi_b-D$
whereas in the strong depletion regime ($\phi_s\ll \phi_b$)

\bea
\phi_s &\simeq &\phi_b\left(\frac{D}{\xi_b}\right)^{5/3}\sim
\phi_b^{9/4}\gamma_1^{-5/2} \nonumber \\
& &\nonumber \\
D&=&a\left(\frac{a^2\gamma_1}{k_B T}\right)^{-3/2} \nonumber \\
& &\nonumber \\
\Gamma &\simeq& -\phi_ba^{-3}(\xi_b-D)\sim \phi_b^{1/4}\nonumber \\
& &\nonumber \\
\gamma-\gamma_0 &\simeq& -\frac{k_B T}{a^2} \phi_b^{3/2}
\label{d27}
\eea

Note that the scaling of the surface excess and surface tension
with the bulk concentration, $\phi_b$ is similar to that obtained
by the mean--field theory approach in Sect. 3.1.2.

\subsection{Proximal Region Corrections}

So far we did not address any corrections in the proximal region:
$a<z<D$ for the many chain adsorption. In the mean--field theory
picture the profile in the proximal region is featureless and
saturates smoothly to its extrapolated surface value, $\phi_s>0$.
However, in relation to surface critical phenomena which is in
particular relevant close to the adsorption-desorption phase
transition (the so-called `special' transition),  the polymer
profile in the proximal region has a scaling form with another
exponent $m$.

\be
\phi(z)\simeq \phi_s\left(\frac{a}{z}\right)^m
\label{d28}
\ee
\noindent where $m=(1-\varphi-\nu)/\nu$ is the proximal exponent,
Eq.~(\ref{proxm}). This is similar to the single chain treatment
in Sect. 2.

For good solvents, one has $m\simeq 1/3$, as was derived  using
analogies with surface critical phenomena, exact enumeration of
polymer configurations, and Monte-Carlo simulations~\cite{ekb82}.
It is different from the exponent 4/3 of the central region.

With the proximal region correction, the polymer profile
can be written as~\cite{dgp83}

\bea
\phi(z) \simeq
     \left\{
     \begin{array}{llll}
     & \phi_s & {\rm for} &  0<z<a  \\
     &&& \\
     &   \phi_s\left(\frac{a}{z}\right)^{1/3} & {\rm for} &  a<z<D  \\
     &&& \\
     & \phi_s\left(\frac{a}{z}\right)^{1/3}\left(\frac{D}{z+D}\right)
                                    & {\rm for} & D<z<\xi_b  \\
                \end{array} \right.
\label{d29}
\eea

\bigskip \noindent where

\medskip
\be
\phi_s=\frac{a}{D}
\label{d30}
\ee

\medskip
\noindent
The complete adsorption profile is shown in Fig.~8a.
By minimization of the free energy with respect to the layer
thickness $D$ it is possible to show that $D$
is proportional to  $1/\gamma_1$

\be
D\sim \gamma_1^{-1}
\label{d31}
\ee

\noindent
in accord with the exact field-theoretic results for a single
chain as discussed in Sect. 2.

The surface concentration, surface excess and surface tension have
the following scaling \cite{dgp83}:

\bea
\phi_s &\simeq& \frac{a} {D} \sim \gamma_1 \nonumber\\
&&\nonumber\\
\Gamma &\simeq& a^{-3}D \left(\frac{a}{D}\right)^{4/3} \sim \gamma_1^{1/3}
\nonumber\\
&&\nonumber\\
\gamma-\gamma_0 &\simeq& -\frac{\gamma_1 a^2}{k_B T}\gamma_1 \sim \gamma_1^2
\label{d32}
\eea

Note the differences in the scaling of the surface tension
and surface excess in Eq.~(\ref{d32}) as compared with the
results obtained with no proximity exponent ($m=0$) in the previous
section, Eq.~(\ref{d25}).

At the end of our discussion of polymer adsorption from solutions,
we would like to add that for the case of adsorption from dilute
solutions, there is an intricate crossover from the single-chain
adsorption behavior, as discussed in Sect. 2, to the adsorption
from semi-dilute polymer solutions, as discussed in this
section~\cite{bd87}. Since the two-dimensional adsorbed layer has
a higher local polymer concentration than the bulk, it is possible
that the adsorbed layer forms a two-dimensional semi-dilute state,
while the bulk is a truly dilute polymer solution. Only for
extremely low bulk concentration or for very weak adsorption
energies  the adsorbed layer has a single-chain structure with no
chain crossings between different polymer chains.

\subsection{Loops  and Tails}
It has been realized quite some time ago that the so-called
central region of an  adsorbed polymer layer is characterized  by
a rather broad distribution of loop and tail
sizes~\cite{fleer,taleoftails,johner93}. A loop is defined as a
chain region located between two points of contact with the
adsorbing surface, and a tail is defined as the chain region
between the free end and the closest contact point to the surface,
while a train denotes a chain section which is tightly bound
to the substrate
(see Fig.~3a). The relative statistical weight of loops and tails
in the adsorbed layer is clearly of importance to applications.
For example, it is expected that polymer loops which are bound at
both ends to the substrate are more prone to entanglements with
free polymers than tails and, thus, lead to enhanced friction
effects. It was found in detailed numerical mean--field theory
calculations that the external part of the adsorbed layer is
dominated by dangling tails, while the inner part  is mostly built
up by loops~\cite{fleer,taleoftails}.

Recently, an analytical theory was formulated which correctly
takes into account the separate contributions of loops and tails
and which thus goes beyond the {\em ground state dominance}
assumption made in ordinary mean--field theories.  The theory
predicts  that a crossover between tail-dominated and
loop-dominated regions occurs at some distance $z^*\simeq a
N^{1/(d-1)}$~\cite{semenov95} from the surface, where $d$ is the
dimension of the embedding space. It is well known that mean--field
theory behavior can formally be obtained by setting the embedding
dimensionality equal to the upper critical dimension, which is for
self-avoiding polymers given by $d=4$ ~\cite{Cloizeaux}. Hence,
the above expression predicts a crossover in the adsorption
behavior at a distance $z^*\simeq a N^{1/3}$. For good-solvent
conditions in three dimensions ($d=3$), $z^*\simeq a N^{1/2}$. In
both cases, the crossover occurs at a separation much smaller than
the size of a free polymer $R_e\sim a N^{\nu}$ where, according to
the classical Flory argument \cite{Flory1}, $\nu = 3/(d+2)$.

A further rather subtle result of these improved mean--field
theories is the occurrence of a depletion hole, \ie, a region at a
certain separation from the adsorbing  surface where the monomer
concentration is smaller than the bulk concentration
\cite{semenov95}. This depletion hole results from an interplay
between the depletion of free polymers from the adsorbed layer and
the slowly decaying density profile due to dangling tails. It
occurs at a distance from the surface comparable with the radius
of gyration of a free polymer, but also shows some dependence on
the bulk polymer concentration. These and other effects, related
to the occurrence of loops and tails in the adsorbed layer, have
been recently reviewed \cite{semrev}.

\section{Interaction between Two Adsorbed Layers}
\setcounter{equation}{0}

One of the many applications of polymers lies in their influence
on the behavior of colloidal particles suspended in a
solvent~\cite{napper}. If the polymers do not adsorb on the
surface of the colloidal particles but are repelled from it, a
strong  attraction between the colloidal particles results from
this polymer--particle depletion, and can lead to polymer-induced
flocculation~\cite{jldg79}. If the polymers adsorb uniformly on
the colloidal surface (and under good-solvent conditions), they
show the experimentally well-known tendency to stabilize colloids
against flocculation, \ie, to hinder the colloidal particles from
coming so close that van-der-Waals attractions will induce
binding.  We should also mention that in other applications,
adsorbing high-molecular weight polymers are used in the opposite
sense as flocculants to induce binding between unwanted sub-micron
particles and, thereby, removing them from the solution. It
follows that adsorbing polymers can have different effects on the
stability of colloidal particles, depending on the detailed
parameters.

Hereafter, we assume the polymers to
form an adsorbed layer around the colloidal particles, with a
typical thickness much smaller than the particle radius, such that
curvature effects can be neglected. In that case, the effective
interaction between the colloidal particles with adsorbed polymer
layers can be traced back to the interaction energy between two
planar substrates covered with polymer adsorption layers.
In the case
when  the approach of the two particles is slow and the adsorbed
polymers are in {\em full equilibrium} with the polymers in  solution,
the interaction between two opposing adsorbed layers is predominantly
attractive~\cite{dg82,scheut1}, mainly
because polymers form bridges between the two surfaces.
Recently, it has been shown that there is
a small repulsive component
to the interaction at large separations~\cite{avalos}.

The typical equilibration times of polymers are extremely long.
This holds in particular for adsorption and desorption processes,
and is due to the slow diffusion of polymers and their rather high
adsorption energies. Note that the adsorption energy of a polymer
can be much higher than $k_BT$ even if the adsorption energy of a
single monomer is small since there are typically many monomers of
a single chain attached to the surface. Therefore, for the typical
time scales of colloid contacts, the adsorbed polymers are not in
equilibrium with the polymer solution. This is also true for most
of the experiments done with a surface-force apparatus, where two
polymer layers adsorbed on crossed mica cylinders are brought in
contact.

In all these cases one has a {\em constrained equilibrium}
situation, where the polymer configurations and thus the density
profile can adjust only with the  constraint that the total
adsorbed polymer excess stays constant. This case has been first
considered by de Gennes~\cite{dg82} and he found that two fully
saturated adsorbed layers will strongly repel each other if the
total adsorbed amount of polymer is not allowed to decrease. The
repulsion is mostly due to osmotic pressure and originates from
the steric interaction between the two opposing adsorption layers.
It was experimentally verified in a series of force-microscope
experiments on polyethylene-oxide layers in water (which is a good
solvent for PEO)~\cite{klein82}.

In other experiments, the formation of the adsorption  layer is
stopped before the layer is fully saturated. The resulting
adsorption layer is called {\em undersaturated}. If two of those
undersaturated adsorption layers approach each other, a strong
attraction develops, which only at smaller separation changes to
an osmotic repulsion~\cite{klein84}. The theory developed for such
non-equilibrium conditions predicts that any surface excess lower
than  the one corresponding to full equilibrium will lead to
attraction at large separations~\cite{rossi}. Similar mechanisms
are also at work in colloidal suspensions, if the total surface
available for polymer adsorption is large compared to the total
polymer added to the solution. In this case, the adsorption layers
are also undersaturated, and the resulting attraction is utilized
in the application of polymers as flocculation
agents~\cite{napper}.

A distinct mechanism which also leads to attractive forces between
adsorption layers was investigated in experiments with
dilute polymer solutions  in bad solvents. An example is
given by polystyrene in
cyclohexane below the theta temperature~\cite{klein80}.
The subsequently developed theory~\cite{kleinpincus}
showed that the adsorption layers attract each other
since the local concentration in the outer part of the
adsorption layers is enhanced
over the dilute solution and lies in the unstable
two-phase region of the bulk phase diagram.
Similar experiments were also done at the theta temperature~\cite{ikp90}.

The force apparatus was also used to measure the interaction
between depletion layers~\cite{lk85}, as realized with polystyrene
in toluene, which is a good solvent for  polystyrene but does not
favor the adsorption of PS on mica. Surprisingly, the resultant
depletion force is too weak to be detected.

The various regimes and effects obtained for the interaction
of polymer solutions between two surfaces have recently
been reviewed~\cite{kleinreview}. It transpires that force-microscope
experiments done on adsorbed polymer layers  form an ideal tool
for investigating the basic mechanisms of polymer adsorption,
colloidal stabilization and flocculation.

\section{Adsorption of Polyelectrolytes}
\label{sectionpe}
\setcounter{equation}{0}

Adsorption of charged chains (polyelectrolytes) onto charged
surfaces is a difficult problem, which is only partially
understood from a fundamental point of view. This is the case in
spite of the prime importance of polyelectrolyte adsorption in
many applications \cite{fleer}. We comment here  briefly on the
additional features that are characteristic for the adsorption of
charged polymers on surfaces.

A polyelectrolyte is a polymer where a fraction $f$ of its
monomers are charged. When the fraction is small, $f\ll 1$, the
polyelectrolyte is weakly charged, whereas when $f$ is close to
unity, the polyelectrolyte is strongly charged. There are two
common ways to control $f$~\cite{RPA}. One
way is to polymerize a
heteropolymer using charged and neutral monomers as building
blocks. The charge distribution along the chain is quenched
(``frozen'') during the polymerization stage, and it is
characterized by the fraction of charged monomers on the chain,
$f$. In the second way, the polyelectrolyte is a weak polyacid or
polybase. The effective charge of each monomer is controlled by
the pH of the solution. Moreover, this annealed fraction depends
on the local electric potential. This is in particular important
to adsorption processes since the local electric field close to a
strongly charged surface can be very different from its value in
the bulk solution.

Electrostatic interactions play a crucial role in the adsorption
of polyelectrolytes \cite{fleer,cs88,csf91}. Besides the fraction
$f$ of charged monomers, the important parameters are the surface
charge density (or surface potential in case of conducting
surfaces), the amount of salt (ionic strength of low molecular
weight electrolyte) in solution and, in some cases, the solution
pH. For polyelectrolytes the electrostatic interactions between
the monomers themselves (same charges) are always repulsive,
leading to an effective stiffening of the
chain~\cite{barrat,NetzHenri}. Hence, this interaction will {\em
favor} the adsorption of single polymer chains, since their
configurations are already rather extended~\cite{Netz4}, but it
will {\em oppose} the formation of dense adsorption layers close
to the surface \cite{bao98}. A special case is that of {\em
polyampholytes}, where the charge groups on the chain can be
positive as well as negative resulting in a complicated interplay
of attraction and repulsion between the
monomers~\cite{polyam1,polyam2}. If the polyelectrolyte chains and
the surface are oppositely charged, the electrostatic interactions
between them will {\em enhance} the adsorption.

The role of the salt can be conveniently expressed in
terms of
the Debye-H\"uckel screening length defined as:

\be
\lambda_{\rm DH}=\left(\frac{8\pi c_{\rm salt} e^2}{\varepsilon
k_B T}\right)^{-1/2} \label{f1} \ee

\noindent where $c_{\rm salt}$ is the concentration of monovalent
salt ions, $e$ the electronic charge and $\varepsilon\simeq 80$
the dielectric constant of the water. Qualitatively, the presence
of small positive and negative ions at thermodynamical equilibrium
screens the $r^{-1}$ electrostatic potential at distances
$r>\lambda_{\rm DH}$, and roughly changes its form to
$r^{-1}\exp(-r/\lambda_{\rm DH})$. For polyelectrolyte adsorption,
the presence of salt has a complex effect. It simultaneously
screens the monomer-monomer repulsive interactions as well as the
attractive interactions between the oppositely charged surface and
polymer.

Two limiting adsorbing cases can be discussed separately: (i) a
non-charged surface on which the chains like to adsorb. Here the
interaction between the surface and the chain does not have an
electrostatic component. However, as the salt screens the
monomer-monomer electrostatic repulsion, it leads to enhancement
of the adsorption. (ii) The surface is charged but does not
interact with the polymer besides the electrostatic interaction.
This is called the pure electro-sorption case. At low salt
concentration, the polymer charge completely compensates the
surface charge. At high salt concentration some of the
compensation is done by the salt, leading to a decrease in the
amount of adsorbed polymer.

In practice, electrostatic and other types of interactions with
the surface can occur in parallel, making the analysis more
complex. An interesting phenomenon of {\it over-compensation} of
surface charges by the polyelectrolyte chains is observed, where
the polyelectrolyte chains form a condensed layer and reverse the
sign of the total surface charge. This is used, \eg, to build a
multilayered structure of cationic and anionic polyelectrolytes
--- a process that can be continued for few dozen or even few
hundred times \cite{decher}-\cite{caruso}. 
The phenomenon of over-compensation
is discussed in Refs.~\cite{bao98,joanny_comp} but is still not
very well understood.

Adsorption of polyelectrolytes from semi-dilute solutions is
treated either in terms of a discrete multi-Stren layer model
\cite{fleer,wagen,linse96} or in a continuum approach
\cite{bao98,muthu87,varoqui}. In the latter, the concentration of
polyelectrolytes as well as the electric potential close to the
substrate are considered as continuous functions. Both the polymer
chains and the electrostatic degrees of freedom are treated on a
mean--field theory level. In some cases the salt concentration is
considered explicitly \cite{bao98,varoqui}, while in other works,
(\eg, in Ref.~\cite{muthu87}) it induces a screened Coulombic
interaction between the monomers and the substrate.

In a recent work~\cite{bao98}, a simple theory has
been proposed to treat polyelectrolyte adsorption
from a semi-dilute bulk. The surface was treated as a
surface with constant electric potential. (Note that in other works,
the surface is considered to have a constant charge density.) In
addition, the substrate is assumed to be impenetrable  by the
requirement that the polymer concentration at the wall is zero.

Within a mean--field theory it is possible to write down the
coupled profile  equations of the polyelectrolyte concentration
and electric field, close to the surface, assuming that the small
counterions (and salt) concentration obeys a Boltzmann
distribution. From numerical solutions of the profile equations as
well as scaling arguments the following picture emerges. For very
low salt concentration, the surface excess of the polymers
$\Gamma$ and the adsorbed layer thickness $D$ are decreasing
functions of $f$: $\Gamma\sim D \sim f^{-1/2}$. This effect arises
from a delicate competition  between an enhanced attraction to the
substrate, on one hand, and an enhanced electrostatic repulsion
between monomers, on the other hand.

Added salt will screen both the electrostatic repulsion between
monomers and the attraction to the surface. In presence of salt,
for low $f$, $\Gamma$ scales like $f/c_{\rm salt}^{1/2}$ till it
reaches a maximum value at $f^* \sim (c_{\rm salt} v)^{1/2}$, $v$
being the excluded volume parameter of the monomers. At this
special value, $f=f^*$, the electrostatic contribution to the
monomer-monomer excluded volume $v_{\rm el} \sim f^2\lambda_{\rm
DH}^2$ is exactly equal to the non-electrostatic $v$. For $f>f^*$,
$v_{\rm el} > v$ and the surface excess is a descending function
of $f$, because of the dominance of monomer-monomer electrostatic
repulsion. It scales as $c_{\rm salt}^{1/2}/f$. Chapter 7 of
Ref.~\cite{fleer} contains a fair amount of experimental results
on polyelectrolyte adsorption.

\section{Polymer Adsorption on Heterogeneous Surfaces}
\setcounter{equation}{0}

Polymer adsorption can be coupled in a subtle way with lateral
changes in the chemical composition or density of the surface.
Such a surface undergoing  lateral rearrangements at
thermodynamical equilibrium is called an {\em annealed} surface
\cite{dg90,aj91}. A Langmuir monolayer of insoluble surfactant
monolayers at the air/water interface is an example of such an
annealed surface. As function of the temperature change, a Langmuir
monolayer can undergo a phase transition from a high-temperature
homogeneous state to a low-temperature demixed state, where dilute
and dense phases coexist. Alternatively, the transition from a
dilute phase to a dense one may be induced by compressing the
monolayer at constant temperature, in which case the adsorbed
polymer layer contributes to the pressure~\cite{aazpr94}. The
domain boundary between the dilute and dense phases can act as
nucleation site for adsorption of bulky molecules~\cite{nao96}.

The case where the insoluble surfactant monolayer interacts with a
semi-dilute polymer solution solubilized in the water subphase was
considered in some detail. The phase diagrams of the mixed
surfactant/polymer system were investigated within the framework
of mean--field theory~\cite{ca95}. The polymer enhances the
fluctuations of the monolayer and induces an upward shift of the
critical temperature. The critical concentration is increased if
the monomers are more attracted (or at least less repelled) by the
surfactant molecules than by the bare water/air interface. In the
case where the monomers are repelled by the bare interface but
attracted by the surfactant molecules (or vice versa), the phase
diagram may have a triple point. The location of the polymer
desorption transition line (\ie, where the substrate-polymer
interaction  changes from being repulsive to being attractive)
appears to have a big effect on the phase diagram of the
surfactant monolayer.

\section{Polymer Adsorption on Curved Interfaces 
and Fluctuating Membranes}
\setcounter{equation}{0}

The adsorption of polymers on rough substrates is of high interest to
applications. One  example is the reinforcement of rubbers by filler
particles such as carbon black or silica particles~\cite{vilgis}.
Theoretical models considered sinusoidal surfaces~\cite{hone}
and  rough and corrugated
substrates~\cite{blunt,marquesfractal}. In all cases,
enhanced adsorption was found and rationalized in terms of
the excess surface available for adsorption.

The adsorption on macroscopically  curved  bodies leads to
slightly modified adsorption profiles, and also to contribution to
the elastic bending moduli of the adsorbing surfaces. The elastic
energy of liquid-like membrane
can be expressed in terms of two bending moduli, $\kappa$
and $\kappa_G$. The elastic energy (per unit area) is

\be
\frac{\kappa}{2} (c_1+c_2)^2 +\kappa_G c_1 c_2
\ee

\noindent where $\kappa$ and $\kappa_G$ are the elastic bending
modulus and the Gaussian bending modulus, respectively. The
reciprocals of the principle radii of curvature of the surface are
given by $c_1$ and $c_2$. Quite generally, the effective
$\kappa_G$ turns out to be positive and thus favors the formation
of surfaces with negative Gaussian curvature, as for example an
`egg-carton' structure consisting of many  saddles. On the other
hand, the effective $\kappa$ is reduced, leading to a more
deformable and flexible surface due to the adsorbed polymer
layer~\cite{dg90,elasticity}.

Of particular interest is the adsorption of strongly charged
polymers on oppositely charged spheres, because this is a
geometry encountered in many colloidal science applications
and in molecular biology as well~\cite{Goeler}-\cite{Netz5}.

In other  works, the effects of a modified architecture of the
polymers on the adsorption behavior was considered. For example,
the adsorption of star polymers~\cite{star}
and random-copolymers~\cite{random} was considered.

Note that some polymers exhibit a transition into a glassy state
in concentrated adsorbed layers. This glassy state depends on the details
of the molecular interaction, which are not considered here. It should
be kept in mind that such high-concentration effects can slow down the
dynamics of adsorption considerably and will prolong the reach of equilibrium.

\section{Terminally Attached Chains} \label{sectionbrush}
\setcounter{equation}{0}

The discussion so far assumed that all monomers of a polymer are
alike and therefore show the same tendency to adsorb  to the
substrate surface. For industrial and technological applications,
one is often interested  in {\em end-functionalized polymers}.
These are polymers which attach with one end only  to the
substrate, as is depicted in Fig.~3b, while the rest of the
polymer is not particularly attracted to (or even repelled from)
the grafting surface. Hence, it attains a random-coil structure in
the vicinity of the surface. Another possibility of block
copolymer grafting (Fig.~3c) will be briefly discussed below as
well.

The motivation to study such terminally attached polymers lies in
their enhanced power to stabilize particles  and surfaces against
flocculation.
Attaching a polymer by its one end  to the surface opens up a much
more effective route to stable surfaces. Bridging and creation of
polymer loops on the same surface, as encountered
in the case of homopolymer adsorption, do not occur if the main-polymer
section is chosen such that it does not adsorb to the surface.

Experimentally, the end-adsorbed polymer layer can be built in
several different ways, depending on the application in mind. 
First, one of the polymer ends can be {\em
chemically} bound to the grafting surface, leading to a tight and
irreversible attachment~\cite{auroy1} shown schematically in
Fig.~3b. The second possibility consists of  {\em physical}
adsorption of a specialized end-group which favors interaction
with the substrate. For example, polystyrene chains have been used
which contain a zwitterionic end group that adsorbs strongly on
mica sheets~\cite{taunton}.

Physical grafting is also possible with a suitably chosen diblock
copolymer (Fig.~3c), \eg, a PS-PVP diblock in the solvent toluene
at a quartz substrate~\cite{field}. Toluene is a {\em selective
solvent} for this diblock, \ie, the PVP (poly-vinyl-pyridine)
block is strongly adsorbed to the quartz substrate and forms a
collapsed anchor, while the PS (polystyrene) block is under
good-solvent conditions, not adsorbing to the substrate and thus
dangling into the solvent. General adsorption scenarios for
diblock copolymers have been theoretically discussed, both for
selective and non-selective solvents~\cite{marques}. Special
consideration has been given to the case when the asymmetry of the
diblock copolymer, \ie, the length difference between the two
blocks, decreases~\cite{marques}.

Yet another experimental realization of grafted polymer layers is
possible with  diblock copolymers which are anchored at the
liquid-air~\cite{kent} or at a liquid-liquid interface of two
immiscible liquids~\cite{teppner}; this scenario offers the
advantage that the surface pressure can be directly measured. A
well studied example is a diblock copolymer of PS-PEO
(polystyrene/ polyethylene oxide). The PS block is shorter and
functions as the anchor at the air/water interface as it is not
miscible in water. The PEO block is miscible in water but because
of attractive interaction with the air/water interface it forms a
quasi-two dimensional layer at very low surface coverage. As the
pressure  increases and the area per polymer decreases, the PEO
block is expelled from the surface and forms a quasi polymer
brush.

In the following we simplify the discussion by assuming  the
polymers to be irreversibly grafted at one end to the substrate.
Let us consider the good solvent case in the absence of any
polymer attraction to the surface. The important new parameter
that enters the discussion is the grafting density (or area per
chain) $\sigma$, which is the inverse of the average area that is
available for each polymer at the grafting surface. For small
grafting densities, $\sigma < \sigma^*$, the polymers will be far
apart from each other and hardly interact, as schematically shown
in Fig.~9a. The overlap grafting density is $\sigma^* \sim a^{-2}
N^{-6/5}$ for swollen chains, where $N$ is the polymerization
index~\cite{gennes}.

For large grafting densities, $\sigma > \sigma^*$, the chains
overlap. Since we assume the solvent to be good, monomers repel
each other. The lateral separation between the polymer coils is
fixed by the grafting density, so that the polymers stretch away
from the grafting surface in order to avoid each other, as
depicted in Fig.~9b. The resulting structure is called a polymer
`brush', with a vertical height $h$ which greatly exceeds the
unperturbed coil radius~\cite{gennes,alex}. Similar stretched
structures occur in many other situations, such as diblock
copolymer melts in the strong segregation regime, or polymer stars
under good solvent conditions~\cite{halperin92}. The universal
occurrence of stretched polymer configurations in many seemingly
disconnected situations warrants a  detailed discussion of the
effects obtained with such systems.

\subsection{Grafted Polymer Layer: a Mean--Field Theory Description}

The scaling behavior of the polymer height can be analyzed using a
Flory-like mean--field theory, which is a simplified version of the
original Alexander theory~\cite{alex}. The stretching of the chain
leads to an entropic free energy loss of $h^2/(a^2 N)$ per chain,
and the repulsive energy density due to unfavorable
monomer-monomer contacts is proportional to the squared monomer
density times the dimensionless excluded-volume parameter $v$
(introduced in Sect. 3). The free energy per chain is then

\be
\label{flory}
\frac{{\cal F}}{k_BT} = \frac{3 h^2}{2 a^2 N} +
2 a^3 v \left(\frac{\sigma N}{h}
\right)^2 \frac{h}{\sigma}
\ee

\noindent
where the numerical prefactors were chosen for convenience.
The equilibrium height is obtained by minimizing  Eq.~(\ref{flory})
with respect to $h$, and the result is~\cite{alex}

\be
\label{Floryh} h = N \left(2  v a^5 \sigma /3\right)^{1/3} \ee

\noindent The vertical size of the brush scales linearly with the
polymerization index $N$, a clear signature of the strong
stretching of the polymer chains. At the overlap threshold,
$\sigma^* \sim N^{-6/5}$, the height scales as $h \sim N^{3/5}$,
and thus agrees with the scaling of an unperturbed chain radius in
a good solvent, as it should. The simple scaling calculation
predicts the brush height $h$ correctly in the asymptotic limit of
long chains and strong overlap. It has been confirmed by
experiments~\cite{auroy1,taunton,field} and computer
simulations~\cite{cos87,murat}.

The above scaling result assumes that all chains are stretched to
exactly the same height, leading to a step-like shape of the
density profile. Monte-Carlo and numerical mean--field calculations
confirm the general scaling of the brush height, but exhibit a
more rounded monomer density profile which goes continuously to
zero at the outer perimeter~\cite{cos87}.
A big step towards a better understanding of stretched polymer
systems was made by Semenov~\cite{semenov85}, who recognized the
importance of {\em classical paths} for such systems.

The classical polymer path is defined as the path which minimizes
the free energy, for a given start and end position, and thus
corresponds to the most likely path a polymer takes. The name
follows from the analogy with quantum mechanics, where the
classical motion of a particle is given by the quantum path with
maximal probability. Since for strongly stretched polymers the
fluctuations around the classical path are weak, it is expected
that a theory that takes into account only classical paths, is a
good approximation in the strong-stretching limit. To quantify the
stretching of the brush, let us introduce the (dimensionless)
stretching parameter $\beta$, defined as

\be
\beta \equiv N\left(\frac{3 v^2 \sigma^2 a^4 }{2 }\right)^{1/3}
= \frac{3}{2} \left( \frac{h}{a N^{1/2}} \right)^2
\ee

\noindent
where $h \equiv N (2 v \sigma a^5/3)^{1/3}$ is the
brush height according to Alexander's theory, compare
Eq.~(\ref{Floryh}). The parameter $\beta$ is proportional to the
square of the ratio of the Alexander prediction for the brush
height $h$ and the unperturbed chain radius $R_0 \sim a N^{1/2}$,
and, therefore, is a measure of the stretching of the brush.
Constructing a classical theory in the infinite-stretching limit,
defined as the limit $\beta \rightarrow \infty$, it was shown
independently by Milner et al.~\cite{milner88} and Skvortsov et
al. \cite{skvortsov}
that the resulting normalized monomer volume-fraction  profile
only depends on the vertical distance from the grafting surface.
It has in fact a {\em parabolic} profile given by

\be \label{paraprofile}
\phi(z) = \left(\frac{3 \pi}{4}\right)^{2/3} -
\left(\frac{\pi z }{2 h}\right)^2
\ee

\noindent The brush height, \ie, the value of $z$ for which the
monomer density becomes zero, is given by  $z^* = (6/\pi^2)^{1/3}
h$. The parabolic brush profile has subsequently been confirmed in
computer simulations~\cite{cos87,murat} and
experiments~\cite{auroy1} as the limiting density profile in the
strong-stretching limit, and constitutes one of the cornerstones
in this field. Intimately connected with the density profile is
the distribution of {\em polymer end points}, which is non-zero
everywhere inside the brush, in contrast with the original scaling
description leading to Eq.~(\ref{Floryh}).

However, deviations from the parabolic profile become
progressively important as the length of the polymers $N$ or the
grafting density $\sigma$ decreases. In a systematic derivation of
the mean--field theory for  Gaussian brushes~\cite{netzbrush} it
was shown that the mean--field theory is  characterized  by a
single parameter, namely the stretching parameter $\beta$. In the
limit $\beta \rightarrow \infty$, the difference between the
classical approximation and the mean--field theory vanishes, and
one obtains the parabolic density profile. For finite $\beta$ the
full mean--field theory and the classical approximation lead to
different results and  both show deviations from the parabolic
profile.

In Fig.~10 we show the density profiles for four different values
of $\beta$, obtained with the full mean--field
theory~\cite{netzbrush}. The parameter values used are $\beta =
100$ (solid line), $\beta=10$ (broken line), $\beta = 1$ (
dotted-dashed line), and $\beta = 0.1$ (dotted line). For
comparison, we also show the asymptotic result according to
Eq.~(\ref{paraprofile}) as a thick dashed line. In contrast to
earlier numerical implementations~\cite{fleer}, the
self-consistent mean--field equations were solved in the continuum
limit, in which case the results only depend on the single
parameter  $\beta$ and direct comparison with other continuum
theories becomes possible. Already for $\beta = 100$ is the
density profile obtained within mean--Field theory almost
indistinguishable from the parabolic profile denoted by a thick
dashed line.

Experimentally, the highest values of $\beta$ achievable are in
the range of $\beta \simeq 20$, and therefore deviations from the
asymptotic parabolic profile are important. For moderately large
values of $\beta >10$, the classical approximation (not shown
here), derived from the mean--field theory by taking into account
only one polymer path per end-point position, is still a good
approximation, as judged by comparing density profiles obtained
from both theories~\cite{netzbrush}, except very close to the
surface. The classical theory misses completely the depletion
effects at the substrate, which mean--field theory correctly takes
into account. Depletion effects at the substrate lead to a
pronounced density depression close to the  grafting surface, as
is clearly visible in Fig.~10.

A further interesting question concerns the behavior of individual
polymer paths. As we already discussed for the infinite-stretching
theories ($\beta \rightarrow \infty$), there are polymers paths
ending at any distance from the surface. Analyzing the paths of
polymers which end at a common distance from the wall, two rather
unexpected features are obtained: i) free polymer ends are in
general stretched; and, ii) the end-points lying close to the 
substrate are pointing towards the surface
 (such that the polymer paths first
move away from the grafting surface before moving towards the
substrate), and end-points lying  beyond a certain distance from the
substrate point away from the surface
(such that the paths move monotonously towards the
surface).
We should point out that these two features have very recently
been confirmed in molecular-dynamics simulations~\cite{seidel}.
They are not an artifact of the continuous self-consistent theory
used in Ref.~\cite{netzbrush} nor are they due to the neglect of
fluctuations. These are interesting results, especially since it
has been long assumed that free polymer ends are
unstretched, based on the assumption that no
forces act on free polymer ends.

Let us now turn to the thermodynamic behavior of a polymer brush.
Using the Alexander description, we can calculate the
free energy per chain
by putting the result for the optimal brush height, Eq.~(\ref{Floryh}),
into the free-energy expression, Eq.~(\ref{flory}). The result is

\be
{\cal F}/k_BT \sim N \left( v \sigma a^2 \right)^{2/3}
\ee

\noindent In the presence of excluded-volume correlations, \ie,
when the chain overlap is rather moderate, the brush height $h$ is
still correctly predicted by the Alexander calculation, but the
prediction for the free energy is in error. Including correlations
\cite{alex}, the free energy is predicted to scale as ${\cal
F}/k_BT \sim N \sigma^{5/6}$. The osmotic surface pressure $\Pi$
is related to the free energy per chain by

\be
\Pi = \sigma^2 \frac{\partial {\cal F}}{\partial \sigma}
\ee
and should thus scale as $\Pi \sim \sigma^{5/3}$ in the
absence of correlations and as $\Pi \sim \sigma^{11/6}$
in the presence of correlations. However,
these theoretical predictions do not compare well
with experimental results for the surface
pressure of a compressed brush~\cite{kent}.
At current, there is no explanation for this discrepancy.
An alternative theoretical method to study tethered chains is the so-called
single-chain mean--field method~\cite{carignano}, where the statistical
mechanics of a single chain is treated exactly, and the interactions
with the other chains are  taken into account on a mean-field level.
This method is especially useful for short chains, where
fluctuation effects are important, and dense systems,
where excluded volume interactions play a role. The calculated
profiles and brush heights agree very well with experiments and
computer simulations, and moreover explain the pressure isotherms
measured experimentally~\cite{kent} and in molecular-dynamics
simulations~\cite{grest}.

As we described earlier, the main interest in end-adsorbed or
grafted polymer  layers stems from their ability to stabilize
surfaces against van-der-Waals attraction. The force between
colloids with grafted polymers is repulsive if the polymers do not
adsorb on the grafting substrates~\cite{collgraft}. This is in
accord with our discussion of the interaction between adsorption
layers, where attraction was found to be mainly caused by bridging
and creation of polymer loops, which of course is absent for
non-adsorbing brushes. A stringent test of brush theories was
possible with accurate experimental measurements of the repulsive
interaction between two opposing grafted polymer layers using a
surface force apparatus~\cite{taunton}. The resultant force could
be fitted very nicely by the infinite-stretching theory of Milner
et al.~\cite{milner88b}. It was also shown that polydispersity
effects, although  rather small experimentally, have to be taken
into account theoretically in order to obtain a good fit of the
data~\cite{milner89}.

\subsection{Solvent, Substrate and Charge Effects on
Polymer Grafting}

So far we assumed that the polymer grafted layer is in contact
with a good solvent. In this case, the grafted polymers try to
minimize their contacts by stretching out into the solvent. If the
solvent is bad, the monomers try to avoid the solvent by forming a
collapsed brush, the height of which is considerably reduced with
respect to the good-solvent case. It turns out that the collapse
transition, which leads to phase separation in the bulk, is
smeared out for the grafted layer and does not correspond to a
true phase transition~\cite{halperin88}. The height of the
collapsed layer scales linearly in $\sigma N$, which reflects the
constant density within the brush, in agreement with
experiments~\cite{auroy2}. Some interesting effects have been
described theoretically~\cite{marko93} and
experimentally~\cite{auroy2} for brushes in mixtures of good and
bad solvent, which can be rationalized in terms of a partial
solvent demixing.

For a theta solvent ($T=T_\theta$) the relevant interaction is
described by the third-virial coefficient; using a simple
Alexander approach similar to the one leading to
Eq.~(\ref{Floryh}), the brush height is predicted to vary with the
grafting density as $ h \sim \sigma^{1/2}$, in agreement with
computer simulations~\cite{theta}.

Up to now we discussed planar grafting layers. Typically, however,
polymers are grafted to curved surfaces. The first study taking
into account curvature effects of stretched and tethered polymers
was done in the context of star polymers~\cite{daoudcotton}. It
was found that chain tethering in the spherical geometry leads to
a universal density profile, showing a densely packed core, an
intermediate region where correlation effects are negligible and
the density decays as $\phi(r) \sim 1/r$, and an outside region
where correlations are important and the density decays as $\phi
\sim r^{-4/3}$. These considerations were  extended using  the
infinite-stretching theory of Milner et al.~\cite{ball},
self-consistent mean--field theories~\cite{dan},  and
molecular-dynamics simulations~\cite{murat91}. Of particular
interest is the behavior of the bending rigidity of a polymer
brush, which can be calculated from the free energy of a
cylindrical and a spherical brush and forms a conceptually simple
model for the bending rigidity of a lipid
bilayer~\cite{milnerbend}.

A different scenario is obtained with special functionalized
lipids with attached water-soluble polymers. If such lipids are
incorporated  into lipid vesicles, the water-soluble polymers
(typically one uses PEG (poly-ethylene glycol) for its non-toxic
properties) form well-separated mushrooms, or, at higher
concentration of PEG lipid, a dense brush. These modified vesicles
are very interesting in the context of drug delivery, because they
show prolonged circulation times in vivo~\cite{allen}. This is
probably due to a steric serum-protein-binding inhibition due to
the hydrophilic brush coat provided by the PEG lipids. Since the
lipid bilayer is rather flexible and undergoes thermal bending
fluctuations, there is an interesting coupling  between the
polymer density distribution and the membrane shape~\cite{lipopol}.
For non-adsorbing, anchored polymers, the membrane will bend away
from the polymer due to steric repulsion, but for adsorbing
anchored polymer the membrane will bend towards the anchored
polymer~\cite{lipopol}.

The behavior of a polymer brush in contact with a solvent, which
is by itself also a polymer, consisting of chemically identical
but somewhat shorter chains than the brush, had been first
considered by de Gennes~\cite{gennes}. A complete scaling
description has been given only recently~\cite{aubouy}. One
distinguishes different regimes where the polymer solvent is
expelled to various degrees from the brush. A somewhat related
question concerns the behavior of two opposing brushes in a
solvent which consists of a polymer solution~\cite{gast}. Here one
distinguishes a regime where the polymer solution leads to a
strong attraction between the surfaces via the ordinary depletion
interaction (compare to Ref.~\cite{jldg79}), but also a high
polymer concentration regime where the attraction is not  strong
enough to induce colloidal flocculation. This phenomenon is called
colloidal restabilization~\cite{gast}.

Another important extension of the brush theory is obtained with
charged polymers \cite{chargebrush}, showing an interesting
interplay of electrostatic interactions, polymer elasticity, and
monomer monomer repulsion. Considering a mixed brush made of
mutually incompatible grafted chains, a novel transition to a
brush characterized by a lateral composition modulation was
found~\cite{marko91}. Even more complicated spatial structures are
obtained with grafted diblock copolymers~\cite{brown}. Finally, we
would like to mention in passing that  these static brush
phenomena have interesting consequences on dynamic properties of
polymer brushes~\cite{halperin88b}.

\section{Concluding Remarks}
\label{conclusion} \setcounter{equation}{0}

We review simple physical concepts underlying the main theories
which deal with equilibrium and static properties of polymers
adsorbed or grafted to substrates. Most of the review dealt with
somewhat ideal situations: smooth and flat surfaces which are
chemically homogeneous; long and linear homopolymer chains where
chemical properties can be averaged on; simple phenomenological
type of interactions between the monomers and the substrate as
well as between the monomers and the solvent.

Even with all the simplifying assumptions, the emerging physical
picture is quite rich and robust. Adsorption of polymers from dilute
solutions can be understood in terms of
single-chain adsorption on the substrate. Mean--field theory is quite
successful but in some
cases fluctuations in the local monomer concentration play an
important role. Adsorption from more concentrated solutions offers
rather complex and rich density profiles, with several regimes
(proximal, central, distal). Each regime is characterized by a
different physical behavior. We reviewed the principle theories
used to model the polymer behavior. We also mentioned briefly more
recent ideas about the statistics of polymer loops and tails.

The second part of this review is about polymers which are
terminally grafted on one end to the surface and are called
polymer brushes. The theories here are quite different since the
statistics of the grafted layer depends crucially on the fact that
the chain is not attracted to the surface but is forced to be in
contact to the surface since one of its ends is chemically or
physically bonded to the surface. Here as well we review the
classical mean--field theory and more advanced theories giving the
concentration profiles of the entire polymer layer as well as that
of the polymer free ends.

We also discuss additional factors that have an effect on the
polymer adsorption and grafted layers: the quality of the solvent,
undulating and flexible substrates such as fluid/fluid interfaces
or lipid membranes; adsorption and grafted layer of charged
polymers (polyelectrolytes); adsorption and grafting on curved
surfaces such as spherical colloidal particles.

Although our main aim was to review the theoretical progress in
this field, we mention many relevant experiments. In this active
field several advanced experimental techniques are used to probe
adsorbed or grafted polymer layers: neutron scattering, small angle
high-resolution x-ray scattering, light scattering using
fluorescent probes, ellipsometry, surface isotherms as well as
using the surface force apparatus to measure forces between two
surfaces.

The aim of this chapter is to review the wealth of knowledge on
how flexible macromolecules such as linear polymer chains behave
as they are adsorbed or grafted to a surface (like an oxide). This
chapter should be viewed as a general introduction to these
phenomena. Although the chapter does not offer any details about
specific oxide/polymer systems, it can serve as a starting point
to understand more complex systems as encountered in applications
and real-life experiments.

\vspace{3cm}
\newlength{\tmp}
\setlength{\tmp}{\parindent}
\setlength{\parindent}{0pt}
{\em Acknowledgments}
\setlength{\parindent}{\tmp}

We would like to thank I. Borukhov  and H. Diamant for discussions
and comments. One of us (DA) would like to acknowledge partial
support from the Israel Science Foundation founded by the Israel
Academy of Sciences and Humanities --- Centers of Excellence
Program, the Israel--US Binational Science Foundation (BSF) under
grant no. 98-00429 and the Tel Aviv University Basic Research
Fund. This work has been completed while both of us visited the
Institute for Theoretical Physics, UCSB.


\pagebreak
\section*{Figure Captions}

\begin{itemize}

 \item[{\bf Fig.~1}]
 Schematic view of different polymers.
a) Linear homopolymers, which are the main subject of this
chapter. b) Branched polymers. c) Charged polymers or
polyelectrolytes, with a certain fraction of charged groups.

\item[{\bf Fig.~2}]

Schematic profile of the monomer volume fraction $\phi(z)$ as a
function of the distance $z$ from a flat substrate as appropriate
a) for the case of adsorption, where the substrate attracts
monomers, leading to an increase of the polymer concentration
close to the wall; and, b) for the case of depletion, where the
substrate repels the monomers  leading to a depression of the
polymer concentration close to the wall. The symbol $\phi_b$
denotes the bulk volume fraction, \ie, the monomer volume fraction
infinitely far away from the wall, and $\phi_s$ denotes the
surface volume fraction right at the substrate surface.

\item[{\bf Fig.~3}]

The different adsorption mechanisms discussed in this chapter: a)
adsorption of a homopolymer, where each monomer has the same
interaction with the substrate. The `tail', `train' and `loop'
sections of the adsorbing chain are shown; b) grafting of an
end-functionalized polymer via a chemical or a physical bond, and;
c) adsorption of a diblock copolymer where  one of the two block
is attached to the substrate surface, while the other is not.

\item[{\bf Fig.~4}]

Different possibilities of substrates: a) the prototype, a flat,
homogeneous substrate; b) a corrugated, rough substrate. Note that
experimentally, every substrate exhibits some degree of roughness
on some length scale; c) a spherical adsorption substrate, such as
a colloidal particle. If the colloidal radius is much larger than
the polymer size, curvature effects (which means the deviation
from the planar geometry) can be neglected; d) a flat but
chemically heterogeneous substrate.

\item[{\bf Fig.~5}]

a) A polymer chain can be described as a chain of bonds of length
$a$, with fixed torsional angles $\theta$, reflecting the chemical
bond structure,
 but with  freely rotating  rotational
angles; b) the simplified model, appropriate for theoretical
calculations, consists of a structureless line, governed by some
bending rigidity or line tension. This model chain is used when
the relevant length scales are much larger than the monomer size,
$a$.

\item[{\bf Fig.~6}]
A typical surface potential felt by a monomer as a function of the
distance $z$ from an adsorbing wall. First the wall is
impenetrable. Then, the attraction is of strength $U$ and range
$B$. For separations larger than $B$, typically a long-ranged tail
exists and is modeled by $-b z^{-\tau}$.

\item[{\bf Fig.~7}]
Schematic drawing of single-chain adsorption. a) In the limit of
strong coupling, the polymer decorrelates into a whole number of
blobs (shown as dashed circles) and the chain is confined to a
layer thickness $D$, of the same order of magnitude as the
potential range $B$; b) in the case of weak coupling, the width of
the polymer layer $D$ is much larger than the interaction range
$B$ and the polymer forms large blobs, within which the polymer
is not perturbed by the surface.

\item[{\bf Fig.~8}]
a) The schematic density profile for the case of adsorption from a
semidilute solution; we distinguish a layer of molecular thickness
$z \sim a$ where the polymer density depends on details of the
interaction with the substrate and the monomer size, the proximal
region $a < z<  D$ where the decay of the density is governed by a
universal power law (which cannot be obtained within mean--field
theory), the central region for $D< z< \xi_b$ with a self-similar
profile, and the distal region   for $\xi_b < z$, where the
polymer concentration relaxes exponentially to the bulk volume
fraction $\phi_b$. b) The density profile for the case of
depletion, where the concentration decrease close to the wall
$\phi_s$ relaxes to its bulk value $\phi_b$ at a distance of the
order of the bulk correlation length $\xi_b$.

\item[{\bf Fig.~9}]
For grafted chains, one distinguishes a) the mushroom regime,
where the distance between chains, $\sigma ^{-1/2}$, is larger
than the size of a polymer coil, and b) the brush regime, where
the distance between chains is smaller than the unperturbed coil
size. Here, the chains are stretched away from the wall due to
repulsive interactions between monomers. The brush height $h$
scales linearly with the polymerization index, $h \sim N$, and is
thus larger than the unperturbed coil radius $R_e \sim a N^\nu$.

\item[{\bf Fig.~10}]
Results for the density profile of a strongly compressed brush, as
obtained within a mean--field theory calculation. As the
compression increases, described by the stretching parameter
$\beta$, which varies from 0.1 (dots) to 1 (dash-dots), 10
(dashes), and 100 (solid line), the density profile approaches the
parabolic profile (shown as a thick, dashed line) obtained within
a classical-path analysis (adapted from Ref.~\cite{netzbrush}).

\end{itemize}


\pagebreak

\begin{figure}[tbh]
 \epsfxsize=17cm
 \centerline{\vbox{\epsffile{fig1.ps} } }
\end{figure}

\vfill
\pagebreak

\begin{figure}[tbh]
 \epsfxsize=17cm
 \centerline{\vbox{\epsffile{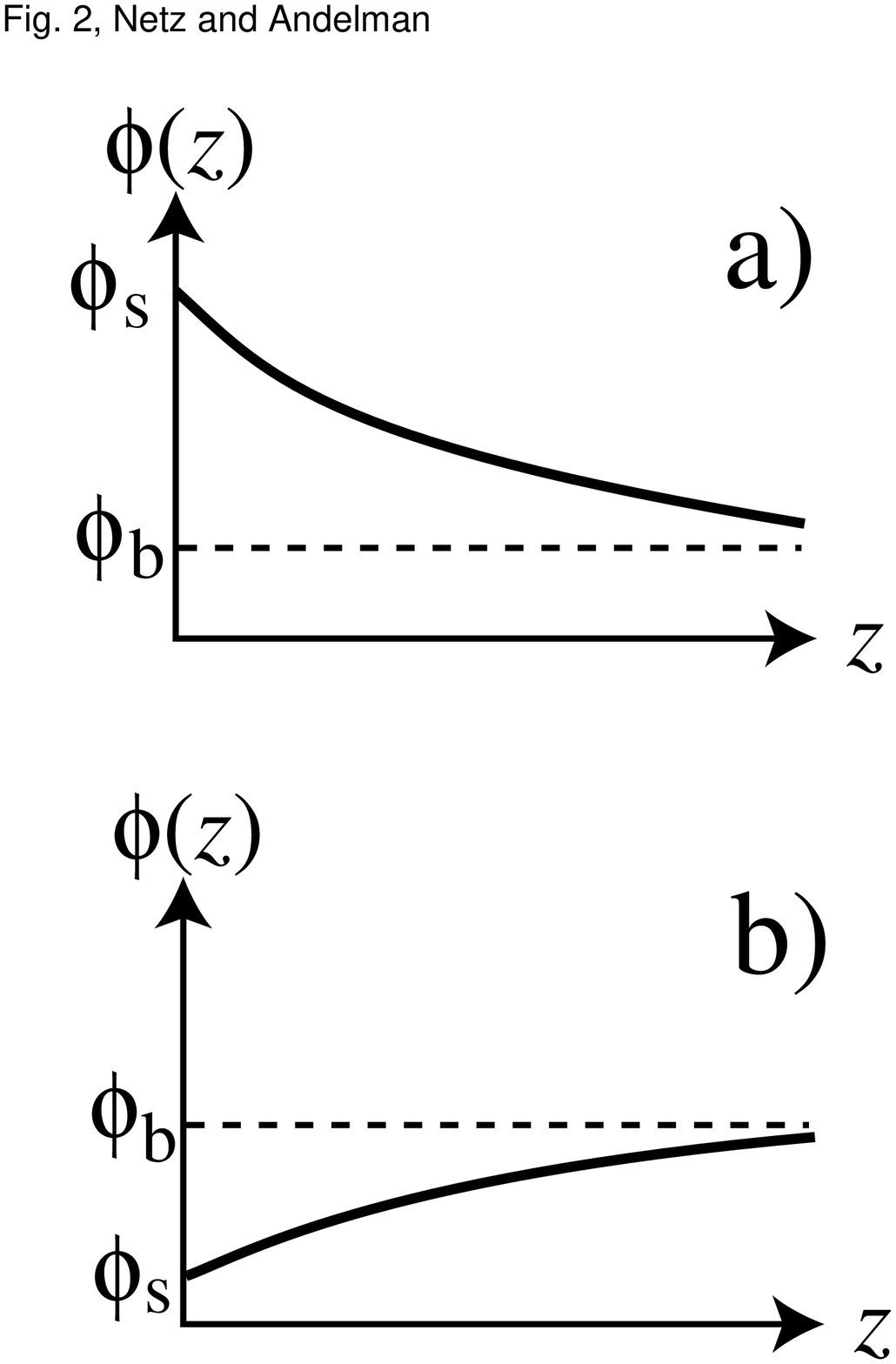} } }
\end{figure}

\vfill
\pagebreak

\begin{figure}[tbh]
 \epsfxsize=17cm
 \centerline{\vbox{\epsffile{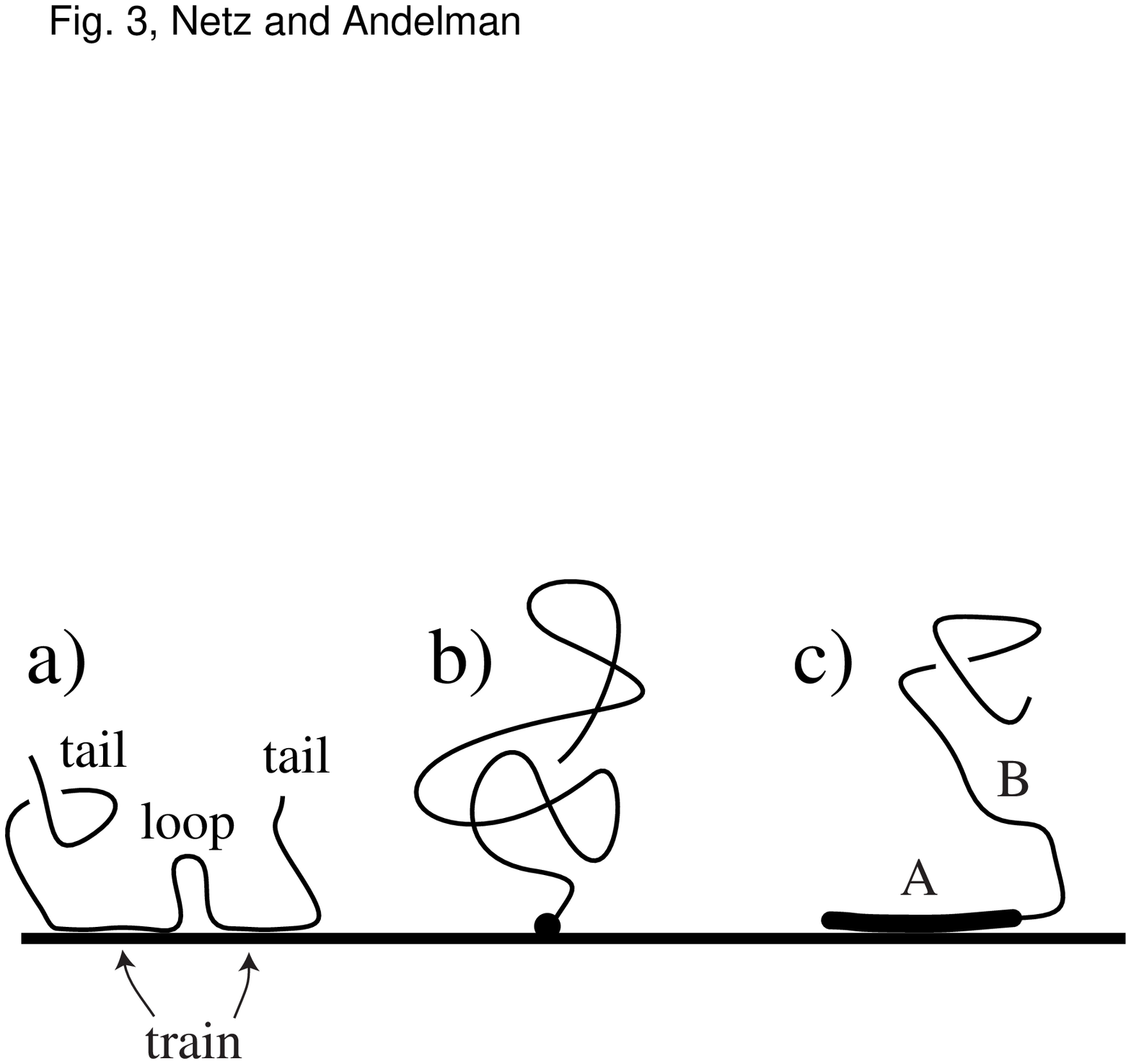} } }
\end{figure}

\vfill
\pagebreak

\begin{figure}[tbh]
 \epsfxsize=17cm
 \centerline{\vbox{\epsffile{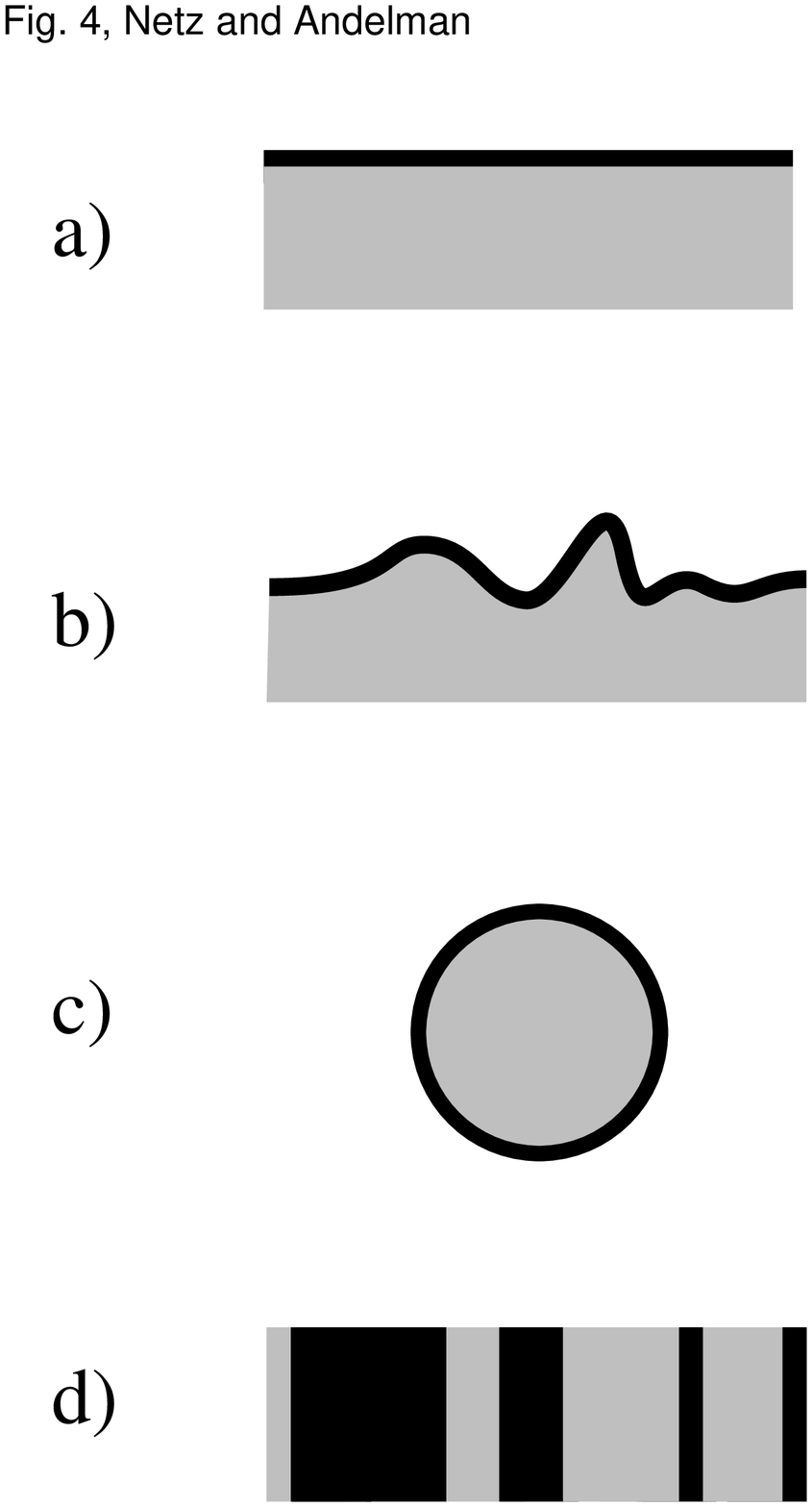} } }
\end{figure}

\vfill
\pagebreak

\begin{figure}[tbh]
 \epsfxsize=17cm
 \centerline{\vbox{\epsffile{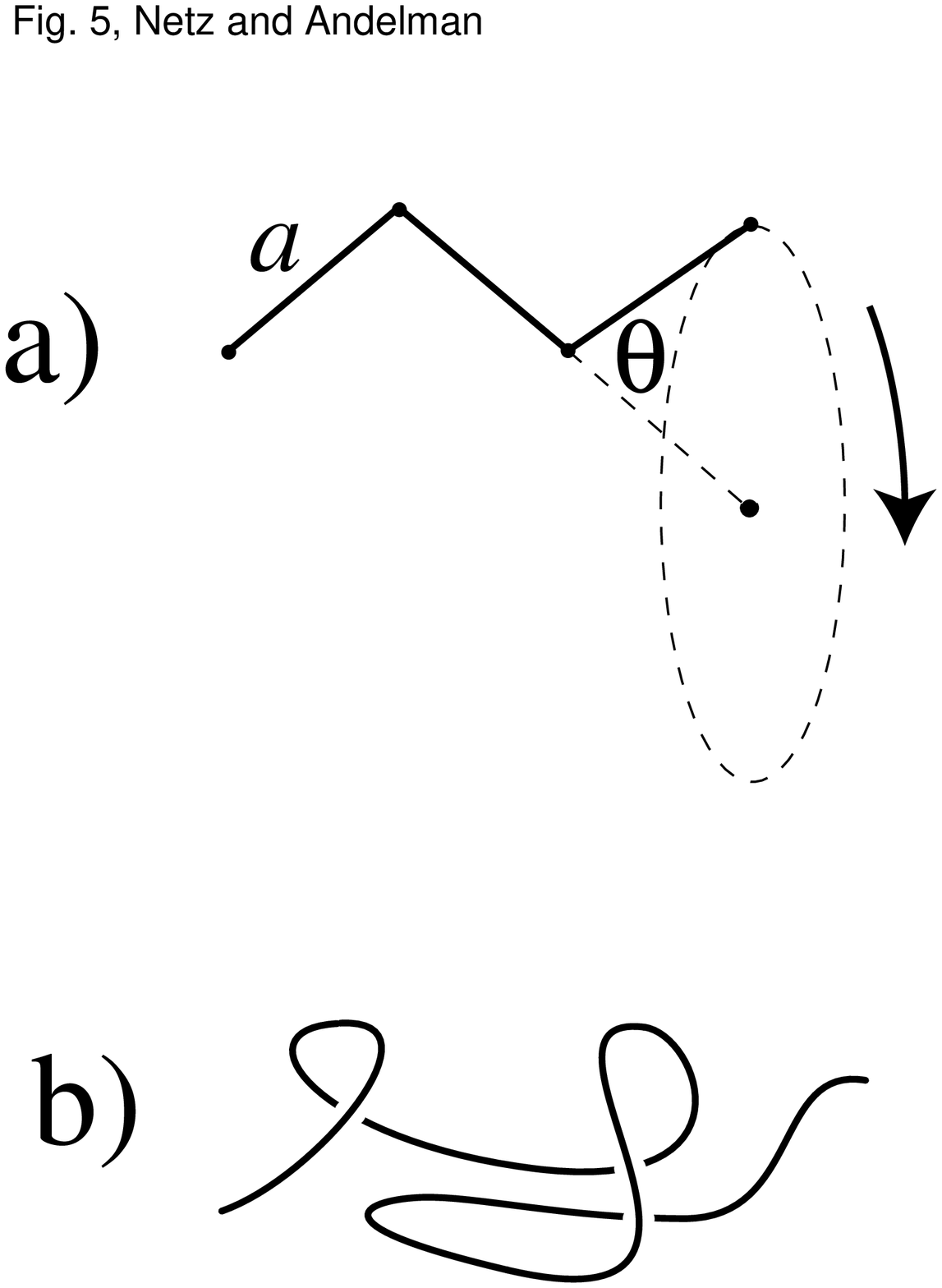} } }
\end{figure}

\vfill
\pagebreak

\begin{figure}[tbh]
 \epsfxsize=17cm
 \centerline{\vbox{\epsffile{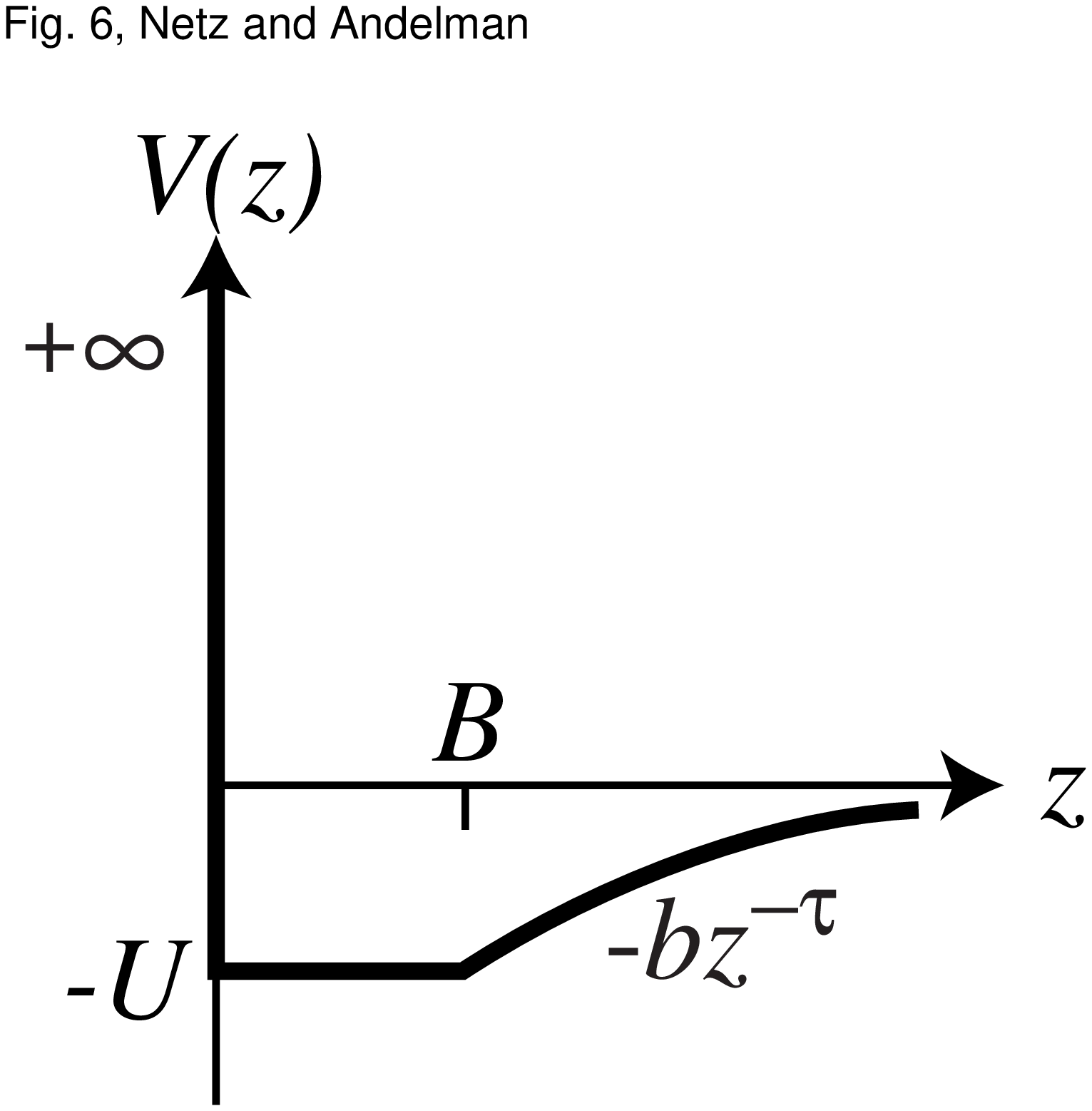} } }
\end{figure}

\vfill
\pagebreak

\begin{figure}[tbh]
 \epsfxsize=17cm
 \centerline{\vbox{\epsffile{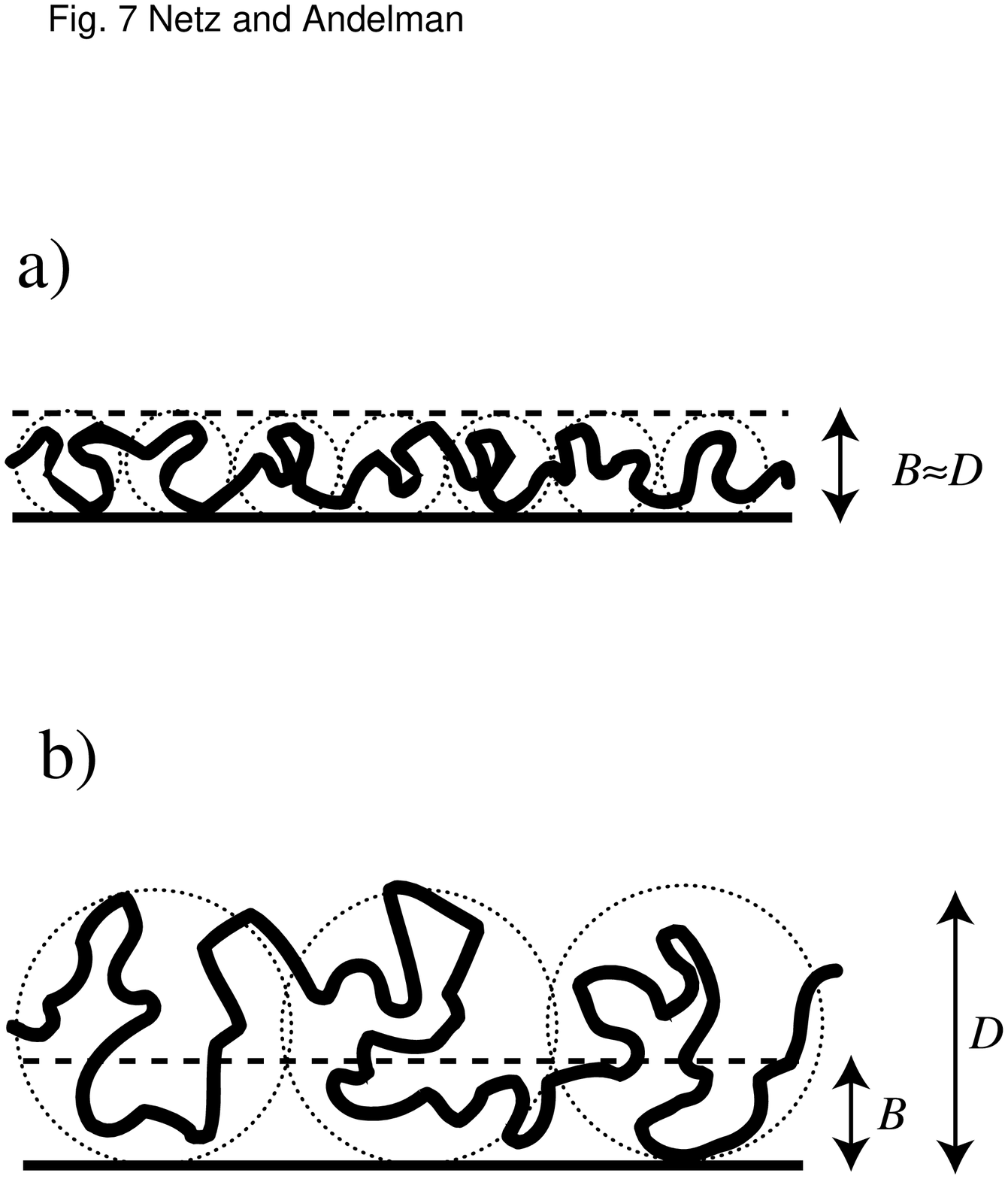} } }
\end{figure}

\vfill
\pagebreak

\begin{figure}[tbh]
 \epsfxsize=17cm
 \centerline{\vbox{\epsffile{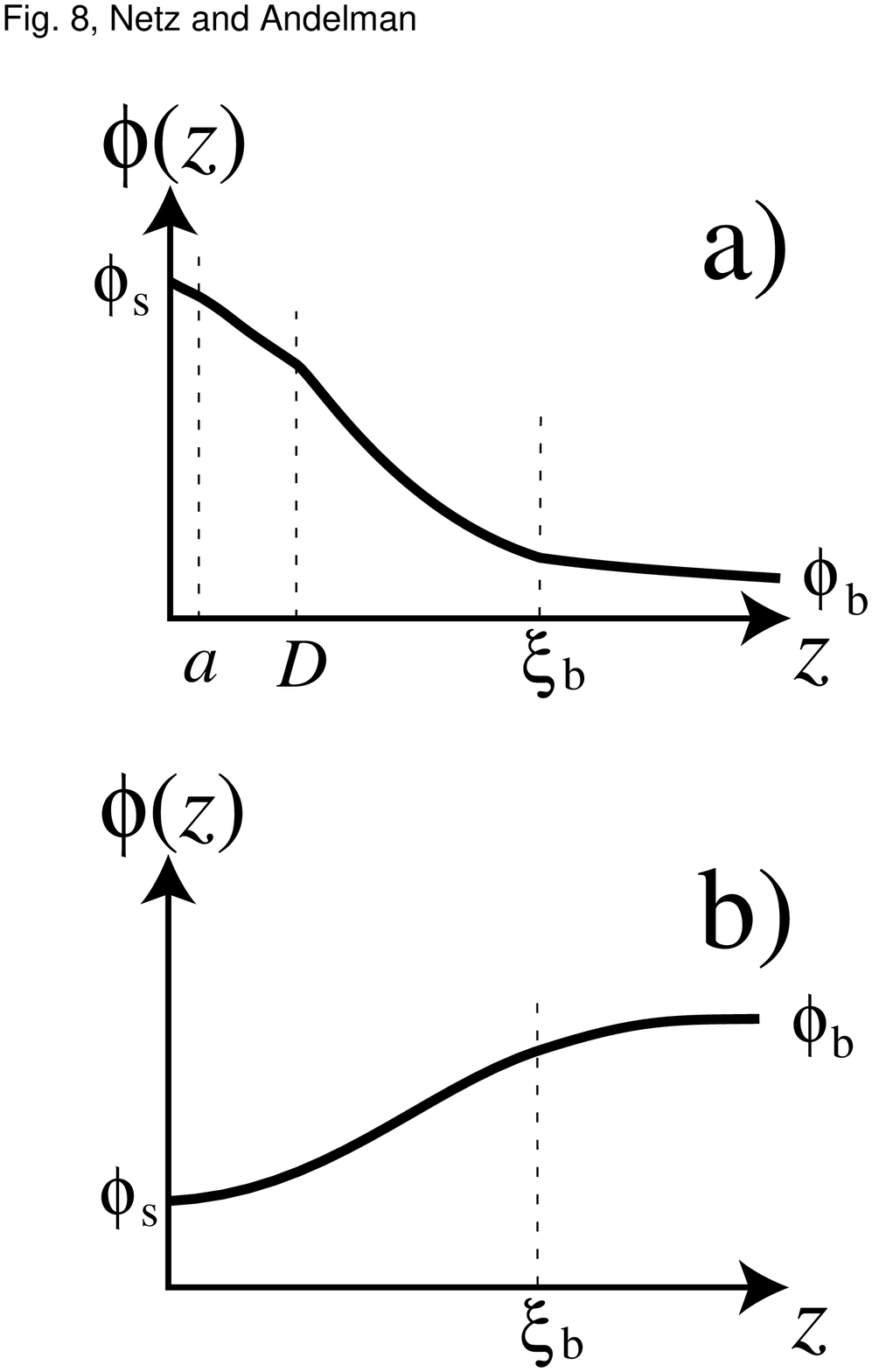} } }
\end{figure}

\vfill
\pagebreak

\begin{figure}[tbh]
 \epsfxsize=17cm
 \centerline{\vbox{\epsffile{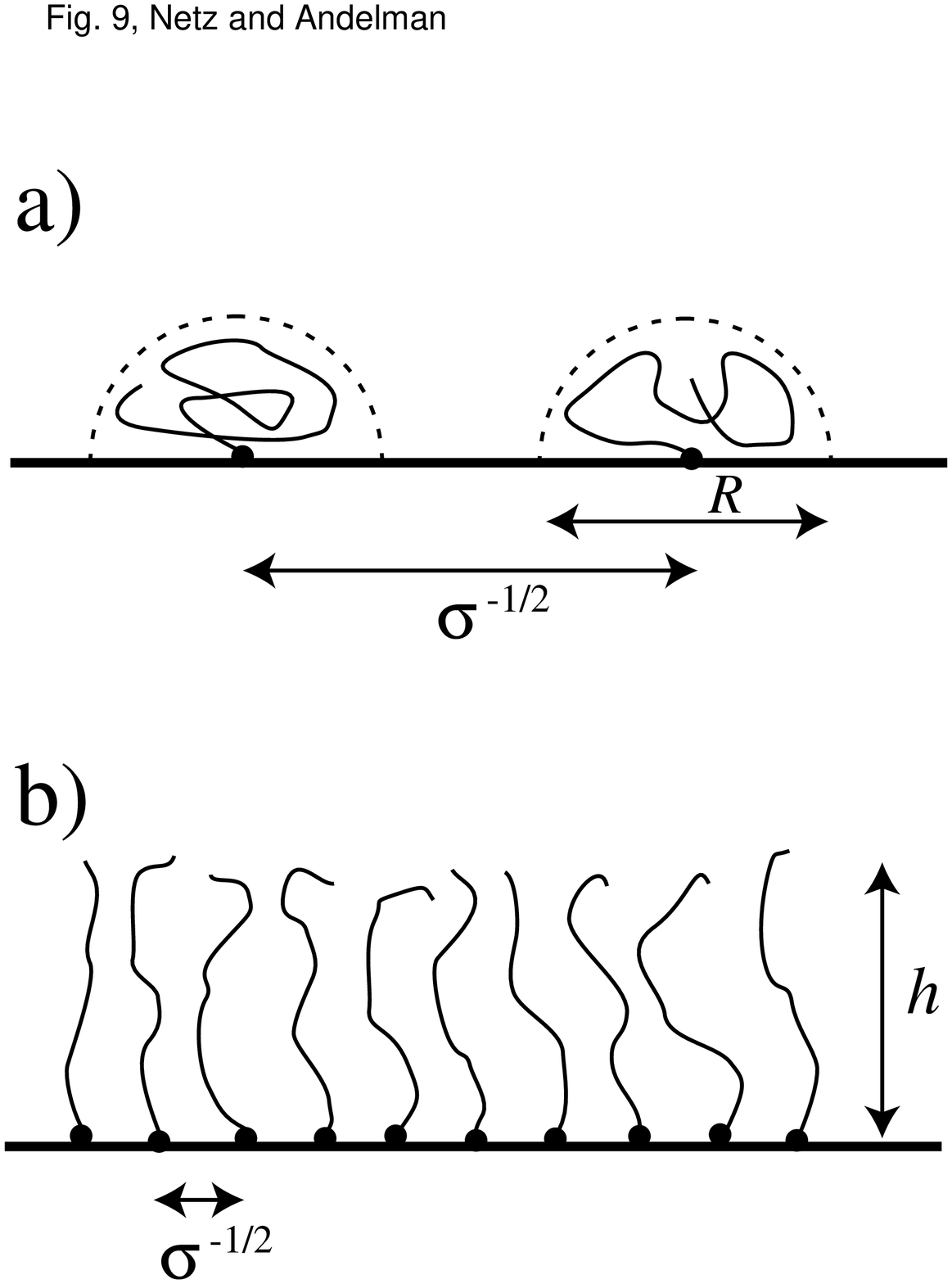} } }
\end{figure}

\vfill
\pagebreak

\begin{figure}[tbh]
 \epsfxsize=17cm
 \centerline{\vbox{\epsffile{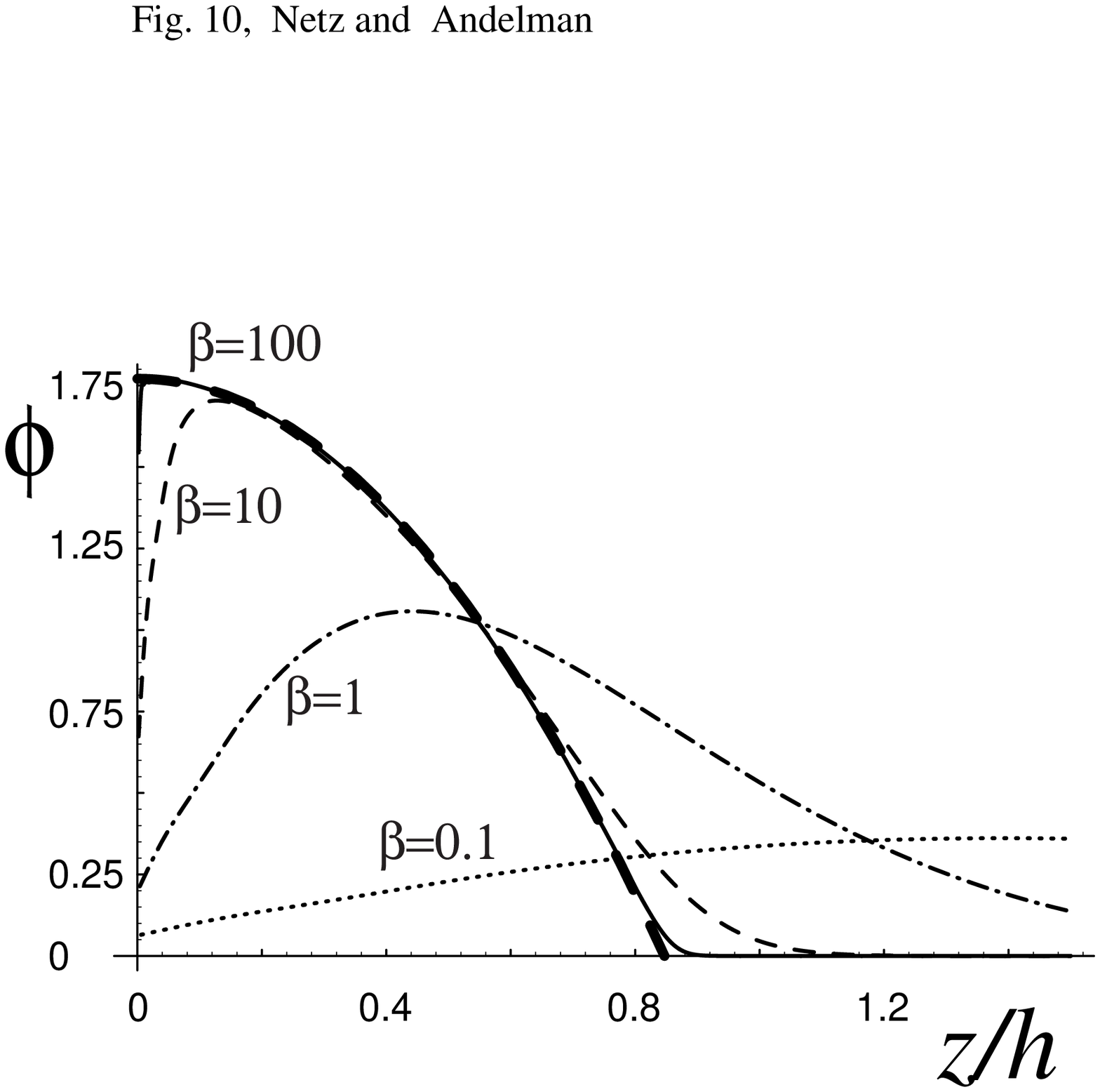} } }
\end{figure}

\vfill



\begin{thebibliography}{99}



\bibitem{cscv86} MA Cohen Stuart, T Cosgrove, B Vincent. Adv Colloid
Interface Sci 24:143-239, 1986.

\bibitem{dg87} PG de Gennes. Adv Colloid Interface Sci 27:189-209, 1987.

\bibitem{s96}I Szleifer. Curr Opin Colloid Interface Sci 1:416-423, 1996.

\bibitem{fl97}GJ Fleer, FAM Leermakers. Curr Opin Colloid Interface
Sci 2:308-314, 1997.

\bibitem{fleer} GJ Fleer, MA Cohen Stuart, JMHM Scheutjens, T Cosgrove,
B Vincent. Polymers at Interfaces. London: Chapman \& Hall, 1993.
%

\bibitem{erich}
E Eisenriegler. Polymers near Surfaces, Singapore: World
Scientific, 1993.

\bibitem{Grosberg}
AYu Grosberg, AR Khokhlov. Statistical Physics of Macromolecules,
New York: AIP Press, 1994.

\bibitem{Flory1}
PJ Flory. Principles of Polymer Chemistry, Ithaca:Cornell
University, 1953.

\bibitem{Yamakawa} H Yamakawa. Modern Theory of Polymer Solutions,
New York: Harper and Row, 1971.

\bibitem{napper}DH Napper. Polymeric Stabilization of Colloidal Dispersions.
London: Academic Press, 1983.

\bibitem{degennes}
PG de Gennes. Scaling Concepts in Polymer Physics. Ithaca: Cornell
University, 1979.

\bibitem{Cloizeaux}
J des Cloizeaux, J Jannink. Polymers in Solution, Oxford:
Oxford University, 1990.


\bibitem{fse53}HL Frish, R Simha, FR Eirish. J Chem Phys 21:365, 1953;
R Simha, HL Frish, FR Eirish. J Phys Chem 57:584, 1953.
%
\bibitem{s62} A Silberberg. J Phys Chem 66:1872, 1962; 66:1884,
1962.
%
\bibitem{e65} SF Edwards. Proc Phys Soc (London) 85:613, 1965;
88:255, 1966.

\bibitem{dm65} EA DiMarzio. J Chem Phys 42:2101, 1965;
EA DiMarzio, FL McCrackin. J Chem Phys 43:539, 1965;
C Hoeve, EA DiMarzio, P Peyser. J Chem Phys 42:2558, 1965.
%
\bibitem{r65} RJ Rubin. J Chem Phys 43:2392, 1965.
%
\bibitem{jr77}IS Jones, P Richmond. J Chem Soc Faraday Trans 2:73, 1977.

\bibitem{Israel}
JN Israelachvili, Intermolecular and Surface Forces, London:
Academic Press, 1992.



\bibitem{daoud77} M Daoud, PG de Gennes. J Phys France 38:85, 1977.

\bibitem{dg69} PG de Gennes. Rep Prog Phys 32:187, 1969.

\bibitem{Flory2}
PJ Flory, Statistical Mechanics of Chain Molecules, Munich: Hanser
Press, 1988.

\bibitem{square}
JMJ van Leeuwen, HJ Hilhorst. Physica A 107:319, 1981; TW
Burkhardt, J Phys A 14:L63, 1981; DM Kroll, Z Phys B 41:345, 1981.

\bibitem{lip89} R Lipowsky, A Baumg\"artner. Phys Rev A 40:2078, 1989;
R Lipowsky. Physica Scripta T29:259, 1989.

\bibitem{netz95}
RR Netz. Phys Rev E 51:2286, 1995.

\bibitem{bor94}
OV Borisov, EB Zhulina, TM Birshtein. J Phys II France 4:913,
1994.

\bibitem{xav98}
X Chatellier, TJ Senden, JF Joanny, JM di Meglio. Europhys
Lett 41:303, 1998.

\bibitem{polyam1}
JF Joanny. J Phys II France 4:1281, 1994; AV Dobrynin, M
Rubinstein,  JF Joanny. Macromolecules 30:4332, 1997.

\bibitem{polyam2}
RR Netz, JF Joanny. Macromolecules 31:5123, 1998.

\bibitem{gennes72}
PG de Gennes. Phys Lett A 38:339, 1972.

\bibitem{ekb82}E Eisenriegler, K Kremer, K Binder. J Chem Phys 77:6296-6320,
1982; E Eisenriegler. J Chem Phys 79:1052-1064, 1983.

\bibitem{dg76}PG de Gennes. J Phys (Paris) 37:1445-1452, 1976.

\bibitem{dgp83} PG de Gennes, P Pincus. J Phys Lett (Paris)
44:L241-L246, 1983.

\bibitem{bd87} E Bouchaud, M Daoud. J Phys (Paris) 48:1991-2000, 1987.



\bibitem{ch58} JW Cahn, JE Hilliard. J Chem Phys 28:258-267, 1958.

\bibitem{dg81}PG de Gennes. Macromolecules 14:1637-1644, 1981.

\bibitem{g92} O Guiselin. Europhys Lett 17:225-230, 1992.

\bibitem{neutscatt}
L Auvray, JP Cotton. Macromolecules 20:202, 1987

\bibitem{neutrefl}
LT Lee, O Guiselin, B Farnoux, A Lapp. Macromolecules 24:2518,
1991; O Guiselin, LT Lee, B Farnoux, A Lapp. J Chem Phys
95:4632, 1991; O Guiselin. Europhys Lett 1:57, 1992.



\bibitem{taleoftails}
JMHM Scheutjens, GJ Fleer,  MA Cohen Stuart. Colloids and
Surfaces 21:285, 1986; GJ Fleer, JMHM Scheutjens, MA Cohen
Stuart. Colloids and Surfaces 31:1, 1988.

\bibitem{johner93}
A Johner, JF Joanny,  M Rubinstein. Europhys Lett 22:591, 1993.

\bibitem{semenov95}
AN Semenov,  JF Joanny. Europhys Lett 29:279, 1995; AN Semenov,
J Bonet-Avalos, A Johner, JF Joanny. Macromolecules 29:2179,
1996; A Johner, J Bonet-Avalos, CC van der Linden, AN Semenov,
JF Joanny. Macromolecules 29:3629, 1996.

\bibitem{semrev}
AN Semenov, JF Joanny, A Johner. In A. Grosberg, ed.
Theoretical and Mathematical Models in Polymer Research, Boston:
Academic Press, 1998.


\bibitem{jldg79} JF Joanny, L Leibler, PG de Gennes. J Polym Sci:
Polym Phys Ed 17:1073-1084, 1979.

\bibitem{dg82}
PG de Gennes. Macromolecules 15:492-500, 1982.

\bibitem{scheut1}
JMHM Scheutjens, GJ Fleer. Macromolecules 18:1882, 1985; GJ
Fleer, JMHM Scheutjens. J Coll Interface Sci 111:504, 1986.

\bibitem{avalos}
J Bonet Avalos, JF Joanny, A. Johner, AN Semenov, Europhys
Lett 35:97, 1996; J Bonet Avalos, A Johner, JF Joanny. J Chem
Phys 101:9181, 1994.

\bibitem{klein82} J Klein, PF Luckham.
Nature 300:429, 1982; Macromolecules 17:1041-1048, 1984.

\bibitem{klein84} J Klein, PF Luckham.
Nature 308:836, 1984; Y Almog, J Klein. J Colloid Interface Sci
106:33, 1985.

\bibitem{rossi} G Rossi,  PA Pincus. Europhys Lett
5:641, 1988;  Macromolecules 22:276-283, 1989.

\bibitem{klein80}
J Klein. Nature 288:248, 1980; J Klein,  PF Luckham.
Macromolecules 19:2007-2010 (1986).

\bibitem{kleinpincus}
J Klein, P Pincus. Macromolecules 15:1129, 1982; K Ingersent, J
Klein,  P Pincus. Macromolecules 19:1374-1381, 1986.

\bibitem{ikp90}  K Ingersent, J Klein, P Pincus. Macromolecules
23:548-560, 1990.

\bibitem{lk85} PF Luckham, J Klein. Macromolecules 18:721-728, 1985.

\bibitem{kleinreview}
J Klein, G Rossi. Macromolecules 31:1979-1988, 1998.





\bibitem{RPA} I Borukhov, D Andelman, H Orland.
Eur Phys J B 5:869-880, 1998.

\bibitem{cs88} MA Cohen Stuart. J Phys (France) 49:1001-1008, 1988.

\bibitem{csf91} MA Cohen Stuart, GJ Fleer, J Lyklema,
W Norde, JMHM Scheutjens. Adv Colloid Interface Sci 34:477-535, 1991.

\bibitem{barrat}
JL Barrat,  Joanny JF {Europhys Lett}  {24}:333, 1993.

\bibitem{NetzHenri}
RR Netz, H Orland. Eur Phys J B 8:81, 1999.

\bibitem{Netz4}
RR Netz, JF Joanny. Macromolecules, 32:9013, 1999.

\bibitem{bao98}I Borukhov, D Andelman, H Orland.
Europhys Lett 32:499, 1995; Macromolecules 31:1665-1671, 1998; 
J Phys Chem B 103:5042-5057, 1999.

\bibitem{decher}
G Decher. Science 277:1232, 1997;
M L\"osche, J Schmitt, G Decher, WG Bouwman,
K Kjaer.   Macromolecules 31:8893, 1998.

\bibitem{donath}
E Donath, GB Sukhorukov, F Caruso,  SA Davis,
H M\"ohwald.  Angew. Chem. Int. Ed. 16:37, 1998; 
GB Sukhorukov, E Donath, SA Davis,
H Lichtenfeld, F Caruso,  VI Popov,
H M\"ohwald.  Polym. Adv. Technol. 9:759, 1998.

\bibitem{caruso}
F Caruso, RA Caruso, H M\"ohwald.
 Science 282:1111, 1998;
F Caruso, K Niikura, DN Furlong,
Y Okahata. Langmuir 13:3422, 1997.

\bibitem{joanny_comp} JF Joanny. Eur Phys J B 9:117, 1999.

\bibitem{wagen}
HA van der Schee, J Lyklema. J Phys Chem 88:6661, 1984;
J Papenhuijzen, HA van der Schee, GJ Fleer. J Colloid Interface Sci
104:540, 1985; OA Evers, GJ Fleer, JHMH Scheutjens, J Lyklema.
J Colloid Interface Sci 111:446, 1985;
HGM van de Steeg, MA Cohen Stuart, A de Keizer, BH Bijsterbosch.
Langmuir 8:8, 1992. 
  
\bibitem{linse96}
P Linse.
  {Macromolecules} {29}:326, 1996.

\bibitem{muthu87}
  M Muthukumar.
  {J Chem Phys} {86}:7230, 1987.

\bibitem{varoqui}
    R Varoqui, A Johner,  A Elaissari.
   {J Chem Phys} {94}:6873, 1991;
   R  Varoqui.
   {J Phys (France) II} {\it 3}:1097, 1993.


\bibitem{dg90}PG de Gennes. J Phys Chem 94:8407, 1990.

\bibitem{aj91}D Andelman, JF Joanny. Macromolecules 24:6040-6041, 1991;
JF Joanny, D Andelman. Makromol Chem Macromol Symp 62:35-41,
1992; D Andelman,  JF Joanny. J Phys II (France) 3:121-138,
1993.

\bibitem{aazpr94} V Aharonson, D Andelman, A Zilman, PA Pincus,
E Rapha\"el. Physica A 204:1-16, 1994; 227:158-160, 1996.

\bibitem{nao96}RR Netz, D Andelman, H Orland. J Phys II (France) 6:1023-1047,
1996.

\bibitem{ca95}X Ch\^atellier, D Andelman. Europhys Lett 32:567-572, 1995;
 X Ch\^atellier, D Andelman. J Phys Chem 22:9444-9455, 1996.

\bibitem{vilgis}
TA Vilgis, G Heinrich. Macromolecules 27:7846, 1994; G Huber, TA Vilgis,
Eur Phys J B 3:217, 1998.

\bibitem{hone}
D Hone, H Ji, PA Pincus. Macromolecules 20:2543, 1987; H Ji, D Hone,
Macromolecules 21:2600, 1988.

\bibitem{blunt}
M Blunt, W Barford, R Ball. Macromolecules 22:1458, 1989.

\bibitem{marquesfractal}
CM Marques, JF Joanny, J Phys France 49:1103, 1988.

\bibitem{elasticity}
JT Brooks, CM Marques,  ME Cates. Europhys Lett 14:713, 1991; J
Phys II France 1:673, 1991; F Clement,  JF Joanny. J Phys II
France 7:973, 1997.

\bibitem{Goeler}
F von Goeler, M Muthukumar. {J Chem Phys} {100}:7796, 1994.

\bibitem{linse}
T Wallin, P Linse. {Langmuir} {12:}305, 1996; {J Phys Chem}
100:17873 1996; {J Phys Chem B} {101}:5506, 1997.

\bibitem{sens}
E Gurovitch, P Sens. {Phys Rev Lett}  {82:}339, 1999.

\bibitem{mateescu}
EM Mateescu, C Jeppesen,  P Pincus. {Europhys Lett} {46:}493,
1999.

\bibitem{Netz5}
RR Netz, JF Joanny. Macromolecules, 32:9026, 1999.


\bibitem{star}
A Halperin, JF Joanny. J Phys II France 1:623, 1991.

\bibitem{random}
CM Marques, JF Joanny. Macromolecules 23:268, 1990; B van Lent, JMHM
Scheutjens. J Phys Chem 94:5033, 1990; JP Donley, GH
Fredrickson. Macromolecules 27:458, 1994.



\bibitem{auroy1}
P Auroy, L Auvray,  L Leger. Phys Rev Lett 66:719, 1991;
Macromolecules 24:2523, 1991; Macromolecules 24:5158, 1991.

\bibitem{taunton}
HJ Taunton, C Toprakcioglu, LJ Fetters,  J Klein. Nature
332:712, 1988; Macromolecules 23:571, 1990.

\bibitem{field}
JB Field, C Toprakcioglu, L Dai, G Hadziioannou, G Smith,  W
Hamilton. J Phys II France 2:2221, 1992.

\bibitem{marques}
CM Marques, JF Joanny,  L Leibler. Macromolecules 21:1051,
1988.
CM Marques, JF Joanny. Macromolecules 22:1454, 1989.

\bibitem{kent}
MS Kent, LT Lee, B Farnoux,  F Rondelez. Macromolecules
25:6240, 1992; MS Kent, LT Lee, BJ Factor, F Rondelez,  GS
Smith. J Chem Phys 103:2320, 1995; HD Bijsterbosch, VO de Haan, AW
de Graaf, M. Mellema, FAM Leermakers, MA Cohen Stuart,  AA van
Well. Langmuir 11:4467, 1995; MC Faur\'e, P Bassereau, MA
Carignano, I Szleifer, Y Gallot,  D Andelman. Euro Phys J B
3:365, 1998.

\bibitem{teppner}
R Teppner, M Harke,  H Motschmann. Rev Sci Inst 68:4177, 1997;
R Teppner H Motschmann. To be published.

\bibitem{gennes}
PG de Gennes. Macromolecules 13:1069, 1980.

\bibitem{alex}
S Alexander. J Phys (France) 38:983, 1977.


\bibitem{halperin92}
A Halperin, M Tirell,  TP Lodge. Adv Pol Sc 100:31, 1992.

\bibitem{cos87}
T Cosgrove, T Heath, B van Lent, F Leermakers,  J Scheutjens.
Macromolecules 20:1692, 1987.

\bibitem{murat}
M Murat, GS Grest. Macromolecules 22:4054, 1989; A Chakrabarti, R
Toral. Macromolecules 23:2016, 1990; PY Lai, K Binder. J
Chem Phys 95:9288, 1991.

\bibitem{semenov85}
AN Semenov. Sov Phys JETP 61:733, 1985.

\bibitem{milner88}
ST Milner, TA Witten,  ME Cates. Europhys Lett 5:413, 1988;
Macromolecules 21:21610, 1988; ST Milner. Science 251:905, 1991.

\bibitem{skvortsov}
AM Skvortsov, IV Pavlushkov, AA Gorbunov, YB Zhulina, OV Borisov,
 VA Pryamitsyn. Polymer Science 30:1706, 1988.

\bibitem{netzbrush}
RR Netz, M Schick. Europhys Lett 38:37, 1997; Macromolecules
31:5105, 1998.

\bibitem{seidel}
C Seidel, RR Netz. Macromolecules, in press.

\bibitem{carignano}
MA Carignano, I Szleifer. J Chem Phys 98:5006, 1993; J Chem
Phys 100:3210, 1994; Macromolecules 28:3197, 1995; J Chem Phys
102:8662, 1995. A detailed summary of tethered layers is given in
I Szleifer, MA Carignano, Advances in Chem Phys XCIV:165,
1996.

\bibitem{grest}
GS Grest. Macromolecules 27:418, 1994.


\bibitem{collgraft}
TA Witten, PA Pincus. Macromolecules 19:2509, 1986; EB Zhulina,
OV Borisov,  VA Priamitsyn. J Coll Surf Sci 137:495, 1990.

\bibitem{milner88b}
ST Milner. Europhys Lett 7:695, 1988.

\bibitem{milner89}
ST Milner, TA Witten,  ME Cates. Macromolecules 22:853, 1989.


\bibitem{halperin88}
A Halperin. J Phys France 49:547, 1988; YB Zhulina, VA Pryamitsyn,
OV Borisov. Polymer Science 31:205, 1989; EB Zhulina, OV
Borisov, VA Pryamitsyn,  TM Birshtein. Macromolecules 24:140,
1991; DRM Williams. J Phys II France 3:1313, 1993.

\bibitem{auroy2}
P Auroy, L Auvray. Macromolecules 25:4134, 1992.

\bibitem{marko93}
JF Marko. Macromolecules 26:313, 1993.

\bibitem{theta}
PY Lai, K Binder. J Chem Phys 97:586, 1992; GS Grest, M
Murat. Macromolecules 26:3108, 1993.

\bibitem{daoudcotton}
M Daoud, JP Cotton. J Phys France 43:531, 1982.

\bibitem{ball}
RC Ball, JF Marko, ST Milner, AT Witten. Macromolecules
24:693, 1991; H Li, TA WItten. Macromolecules 27:449, 1994.

\bibitem{dan}
N Dan, M Tirrell. Macromolecules 25:2890, 1992.

\bibitem{murat91}
M Murat, GS Grest. Macromolecules 24:704, 1991.

\bibitem{milnerbend}
ST Milner, TA Witten.  J Phys France 49:1951, 1988.


\bibitem{allen}
TM Allen, C Hansen, F Martin, C. Redemann,  A Yau-Young.
Biochim Biophys Acta 1006:29, 1991; MJ Parr, SM Ansell, LS Choi,
PR Cullis. Biochim Biophys Acta 1195:21, 1994.

\bibitem{lipopol}
R Lipowsky. Europhys Lett 30:197, 1995; C Hiergeist, R
Lipowsky. J Phys II France 6:1465, 1996; C Hiergeist, VA Indrani,
R Lipowsky. Europhys Lett 36:491, 1996; T Breidenich, RR Netz,
R Lipowsky. Europhys Lett, in press.

\bibitem{aubouy}
M Aubouy, GH Fredrickson, P Pincus,  E Raphael. Macromolecules
28:2979, 1995.

\bibitem{gast}
AP Gast,  L Leibler. Macromolecules 19:686, 1986.

\bibitem{chargebrush}
P Pincus. Macromolecules 24:2912, 1991; OV Porisov, TM Birshtein,
EB Zhulina. J Phys II France 1:521, 1991; F Csaijka, RR Netz,
C Seidel. To be published.

\bibitem{marko91}
JF Marko, TA Witten. Phys Rev Lett 66:1541, 1991.

\bibitem{brown}
G Brown, A Chakrabarti, JF Marko. Macromolecules 28:7817,
1995; EB Zhulina, C Singhm, AC Balazs. Macromolecules 29:8254,
1996.


\bibitem{halperin88b}
A Halperin, S Alexander. Europhys Lett 6:329, 1988; A Johner, JF
Joanny. Macromolecules 23:5299, 1990; C Ligoure, L
Leibler. J Phys France 51:1313, 1990; ST Milner. Macromolecules
25:5487, 1992; A Johner, JF Joanny. J Chem Phys 98:1647, 1993.


\end{thebibliography}
\end{document}